\documentclass[12pt]{iopart}
\usepackage{iopams}  

\usepackage[utf8]{inputenc}
\usepackage[super]{nth}
\usepackage{subfig}
\usepackage{graphicx}
\usepackage{placeins}
\usepackage[margin=1in]{geometry}
\usepackage{amsmath}
\usepackage{appendix}
\usepackage{xcolor}
\usepackage{enumitem}
\usepackage{todonotes}
\usepackage{multicol}

\usepackage[backend=bibtex,sorting=none,style=numeric-comp]{biblatex}
\bibliography{main}

\usepackage{animate}
\usepackage{graphicx}
\usepackage{epstopdf}
\AppendGraphicsExtensions{.gif}

\graphicspath{ {./figs/} }

\usepackage{xspace}

\begin{document}

\title[Smart Pixel Sensors]{Smart pixel sensors: towards on-sensor filtering of pixel clusters with deep learning}

\author{Jieun Yoo$^1$,
Jennet Dickinson$^2$,
Morris Swartz$^3$,
Giuseppe Di Guglielmo$^{2,4}$,
Alice Bean$^5$,
Douglas Berry$^2$, 
Manuel Blanco Valentin$^4$,
Karri DiPetrillo$^6$,
Farah Fahim$^{2,4}$,
Lindsey Gray$^2$,
James Hirschauer$^2$,
Shruti R. Kulkarni$^7$, 
Ron Lipton$^2$,
Petar Maksimovic$^3$,
Corrinne Mills$^1$,
Mark S. Neubauer$^8$,
Benjamin Parpillon$^{2,1}$,  
Gauri Pradhan$^2$,
Chinar Syal$^{2}$, 
Nhan Tran$^{2,4}$, 
Dahai Wen$^{3}$,
Aaron Young$^7$}

\address{$^1$ University of Illinois Chicago, Chicago, IL, 60607, USA}
\address{$^2$ Fermi National Accelerator Laboratory, Batavia, IL 60510, USA}
\address{$^3$ Johns Hopkins University, Baltimore, MD 21218, USA}
\address{$^4$ Northwestern University, Evanston, IL 60208, USA}
\address{$^5$ University of Kansas, Lawrence, KS 66045, USA}
\address{$^6$ The University of Chicago, Chicago, IL 60637, USA}
\address{$^7$ Oak Ridge National Laboratory, Oak Ridge, TN 37831, USA}
\address{$^8$ University of Illinois Urbana-Champaign, Champaign, IL 61801, USA}


\begin{abstract}
Highly granular pixel detectors allow for increasingly precise measurements of charged particle tracks. Next-generation detectors require that pixel sizes will be further reduced, leading to unprecedented data rates exceeding those foreseen at the High Luminosity Large Hadron Collider. Signal processing that handles data incoming at a rate of $\mathcal{O}$(40MHz) and intelligently reduces the data within the pixelated region of the detector \textit{at rate} will enhance physics performance at high luminosity and enable physics analyses that are not currently possible. Using the shape of charge clusters deposited in an array of small pixels, the physical properties of the traversing particle can be extracted with locally customized neural networks. In this first demonstration, we present a neural network that can be embedded into the on-sensor readout and filter out hits from low momentum tracks, reducing the detector's data volume by 54.4-75.4\%. The network is designed and simulated as a custom readout integrated circuit with 28\,nm CMOS technology and is expected to operate at less than 300\,$\mu W$ with an area of less than 0.2\,mm$^2$. The temporal development of charge clusters is investigated to demonstrate possible future performance gains, and there is also a discussion of future algorithmic and technological improvements that could enhance efficiency, data reduction, and power per area. 
\end{abstract}

%
%
%
%
%

\clearpage

\section{Introduction}


High granularity silicon pixel detectors are crucial for disentangling the tremendous numbers of particles produced at high energy colliders. With billions of readout channels and event rates as high as 40\,MHz, these pixel detectors generate petabytes of data per second. To quickly extract the pixel information necessary for high priority physics, we propose to develop intelligent on-chip data reduction. This paper specifically investigates a neural network that can selectively read out pixel clusters based on the incident particle's momentum. 

As the subsystems closest to the interaction point, pixel detectors provide precision spatial measurements that play an essential role in pattern recognition, vertexing, and particle momentum measurements. A traversing charged particle creates electrical signals in a cluster of pixels that, in combination with the pixel sensor's location, provide precise 3D measurements to seed pattern recognition. The pixel size and proximity to the interaction point determine the track impact parameter and momentum resolution. Both the desired spatial precision and the expected density of charged particles are used to decide the pixel dimensions. The current generation ATLAS~\cite{Aad:1129811} and CMS~\cite{Chatrchyan:1129810} experiments at the Large Hadron Collider (LHC) contain pixel detectors with pitches of $50\times250-400\, \mu \textrm{m}^2$ and $100\times 150\, \mu \textrm{m}^2$, respectively, and a thickness of $\mathcal{O}(300\,\mu \textrm{m})$.  At the High Luminosity LHC (HL-LHC), the pixels will be reduced to roughly $50 \times 50\,\mu \textrm{m}^2$ in size and $\mathcal{O}(100\,\mu \textrm{m})$ thick \cite{CERN-LHCC-2017-021,Dominguez:1481838}. 

The particle properties extracted from pixel detector data play a crucial role in physics measurements. In high-luminosity environments, vertex information is essential to distinguish the primary interaction from additional proton-proton interactions occurring in the same bunch crossing (pileup). Impact parameter measurements from the pixel detector also provide key information for reconstructing particles with relatively long lifetimes that decay a measurable distance from the beam axis, including the tau lepton, charm quark, and bottom quark. Correct identification of these particles is necessary for several high-priority searches and measurements, such as characterizing the Higgs boson's couplings to second and third generation fermions. 
In addition, many Beyond the Standard Model scenarios predict new particles that would lead to displaced vertex signatures in the detector. Many hidden valley or dark sector~\cite{strassler2007echoes} models contain signatures with displaced leptons or ``dark QCD'' hadronic showers~\cite{knapen2021perturbative}. Hypothesized new particles decaying to soft, unclustered energy patterns (SUEPs)~\cite{knapen2017triggering} are another distinctive signature, which require pixel information to distinguish it from pileup.

Despite its importance to collider physics programs, pixel detector information is difficult to read out. The large data volume calls for multiple methods of data reduction. Zero suppression is employed to read out only active pixels, and the accumulated charge per pixel is digitized and represented with only a few bits.  Even with these techniques, data rates at the current CMS and ATLAS experiments exceed bandwidth constraints for read out at the collision frequency of 40\,MHz. Collisions of interest are selected using a hardware-based trigger system, which uses information from the other (non-pixel) detector subsystems to select events at a rate of $< 1$\,MHz. Events that contain new physics only in the pixel data are not selected by the low-level trigger and are lost forever.  Furthermore, the more granular pixel detectors built for the HL-LHC and beyond will lead to even higher data rates, which further increases the need for data reduction. 

In this paper, we seek to overcome the limitations of pixel readout with local data reduction in dedicated circuits before transmitting information off of the detector.  While pixel readout has traditionally been treated as a \textit{lossless compression} task, we explore the paradigm of lossy compression within a given event to enable lossless readout over all possible collisions.  To enable this strategy with optimal performance and yet keep the algorithms reconfigurable, we explore the use of neural networks with reprogrammable weights. A machine learning approach is required due to the complicated pulse shapes, a combination of drift and induced currents, generated by the pixel detector. The contribution from those two components is dependent on the sensor geometry where the charged particle impacts the detector and the particle's trajectory. As a first benchmark task, we develop, design, and study a neural network that will filter out pixel clusters originating from low-momentum charged particle tracks.  We then optimize, design, and simulate that neural network in an integrated circuit and study its performance, power, area, and latency.  

The layout of this paper is as follows. The pixel geometry we assume for this study and the simulated pixel sensor data is detailed in Section~\ref{sec:dataset}. Then in Section~\ref{sec:algo}, we introduce the benchmark pixel cluster filtering task, the algorithm designed and optimized for the task, and its performance.  We then detail a preliminary implementation of the algorithm for deployment on-sensor in Section~\ref{sec:implementation}.  Finally in Section~\ref{sec:outlook}, we summarize the results from this study and discuss the additional benchmark tasks and their benefits, as well as ways to further improve the performance and efficiency of the algorithm.

\subsection*{Related Work}

Data compression of silicon tracking information, in both strip and pixel detectors, is explored in Ref.~\cite{Garcia-Sciveres_2014,garcia2014data,garcia2016data,}.  These works contain detailed studies of both lossless and lossy compression of the digital information content of the pixel and strip tracker detectors. These references provide a good baseline for understanding what data compression factor is possible for current tracking systems.   Our work expands on these previous studies by: looking at analog and timing information; considering future pixel dimensions, studying the potential for cluster filtering by the track momentum; and designing a first implementation of on-chip detector algorithms.

Ref.~\cite{fox2021beyond} explores the use of pixel cluster shapes offline (off-detector) to extract direction information and reduce tracking combinatorics and complexity.  The technique described in this study can be used in the future to extract directional information from charge clusters in a single pixel layer, providing similar benefits in terms of reduced algorithm complexity for tracking downstream, which is particularly important for online data processing systems.

Finally, our study relies on previous work for translating neural network algorithms into circuits using the \texttt{hls4ml}~\cite{Duarte:2018ite,fastml_hls4ml} workflow.  In particular, the first implementation using \texttt{hls4ml} to build a reconfigurable ASIC~\cite{di2021reconfigurable} for calorimeter on-detector data compression provides a basis for much of the technology developed in this paper.

\section{Sensor geometry and dataset}
\label{sec:dataset}

\subsection{Simulated data}
\label{sec:tracked}

\begin{figure}[htbp]
  \centering
  \includegraphics[width=0.6\textwidth]{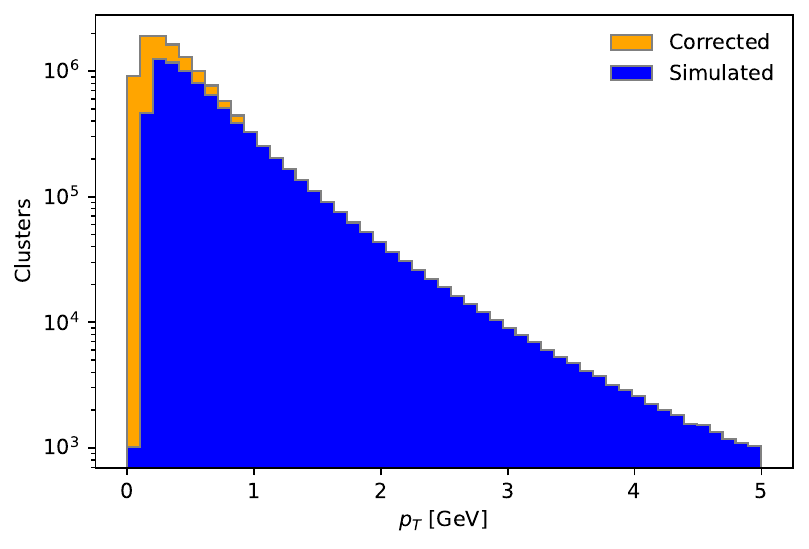}
   \caption{Transverse momentum ($p_T$) of simulated particles (blue) and the $p_T$ distribution corrected for tracking inefficiency (orange).}
  \label{pT}
\end{figure}

The studies in this paper are based on a simulated dataset of silicon pixel clusters produced by charged particles (pions)~\cite{zenodo}.
The kinematic properties of the incident particles are taken from fitted tracks in CMS 13 TeV collision data. Figure \ref{pT} shows the $p_T$ distribution of these particles in blue. Because particles with very low transverse momentum ($p_T$) are not reconstructed as tracks in CMS, the distribution turns on above 100 MeV. The corresponding $p_T$ distribution corrected for losses due to inefficiency of the CMS tracker is shown in orange. 

To study a concrete sensor configuration, we make the following assumptions about our future pixel sensors:
\begin{itemize}
    \item The sensor plane is described by coordinates $x$ and $y$, while the $z$ direction is normal to the sensor. The pixel pitch is taken to be 50\,$\mu m$ $\times$ 12.5\,$\mu m$ in $x\times y$.
    \item The overall pixel sensor area is 16$\times$16\,mm$^{2}$ and its thickness is 100\,$\mu m$.
    \item The sensor is situated on a cylinder of radius 30\,mm, with the particle's origin at its center. 
    Particle interactions are simulated at varying positions of the sensor along the cylinder's axis.
    \item A bias voltage of -100V is applied.
    \item The detector is immersed in a 3.8\,T magnetic field parallel to the $x$ coordinate.
\end{itemize} 
 
The detector response is simulated using a time-sliced version of PixelAV~\cite{pixelav}, which provides: an accurate model of charge deposition by primary hadronic tracks (in particular to model delta rays), a realistic electric field map resulting from the simultaneous solution of Poisson’s Equation, carrier continuity equations, and various charge transport models, an established model of charge drift physics including mobilities, Hall Effect, and 3-D diffusion, a simulation of charge trapping and the signal induced from trapped charge, and a simulation of electronic noise, response, and threshold effects.  A particularly valuable aspect of PixelAV used in this study is time evolution of the drift and induced currents in the pixel sensor.

\begin{figure}[htbp]
  \centering
  \subfloat[]{\includegraphics[width=0.62\textwidth]{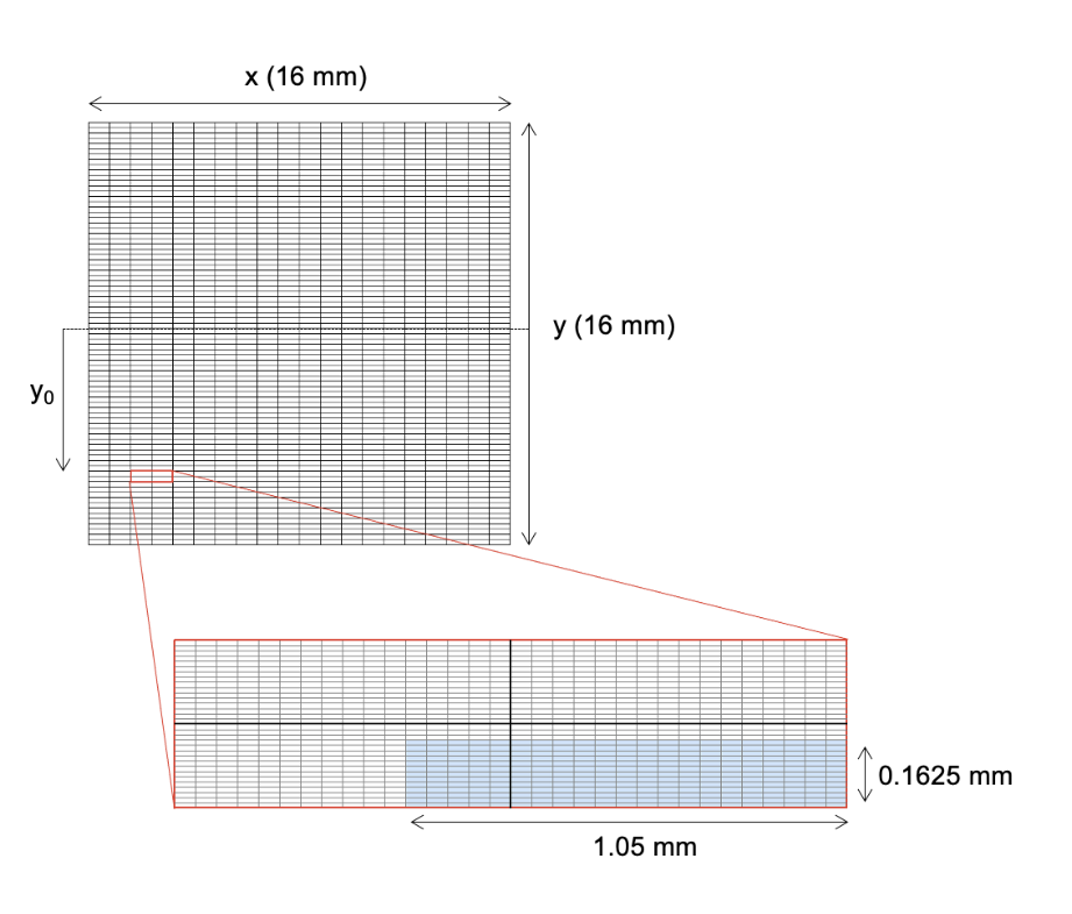}}
  \subfloat[]{\includegraphics[width=0.36\textwidth]{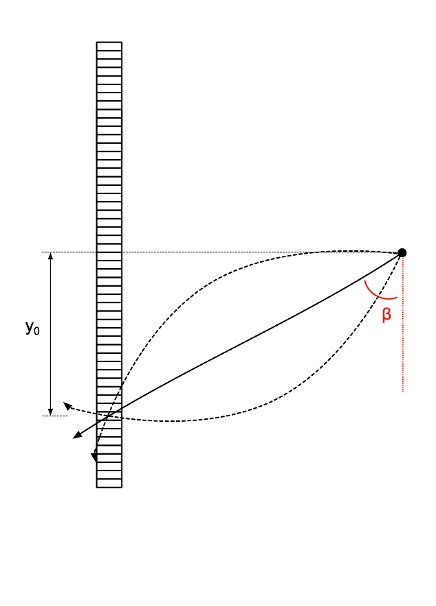}}
   \caption{(a): A schematic of the pixel sensor area and the specific region of interest (blue) of 21$\times$13 pixels for a given cluster.  The magnetic field is parallel to the sensor x coordinate. (b): A diagram of three charged particles traversing our simulated silicon sensor at the same $y_0$ position. The sensor is viewed in the bending plane of the magnetic field. The solid track corresponds to a charged particle with high $p_T$, while the two dashed tracks correspond to low $p_T$ particles with opposite charge.}
  \label{sensor-cartoon}
\end{figure}

Figure \ref{sensor-cartoon} sketches out key features of the pixel sensor and corresponding strategies employed by this paper.  Within the pixel sensor area, we define a \textit{cluster region of interest}, shown in blue, which corresponds to 21$\times$13 pixels in $x$ and $y$, respectively. This region is large enough to fully encompass a charge cluster and serves as input to the ML algorithm used to extract cluster features. The position $(x,y)$ where the charged particle traverses the sensor mid-plane is uniformly distributed across the central $3\times3$ pixel array. The shape of the charge deposited in the pixel array is sensitive to this position and to the particle's angle of incidence. The incident angle in the $x-z$ plane is denoted by $\alpha$, and by $\beta$ in the $y-z$ plane. Due to the bending of charged particle tracks in the magnetic field, the shape of the charge cluster also depends on the particle's $p_T$, which is highly correlated with $\beta$.  The shape of the cluster also depends strongly on its azimuthal position with respect to the center of the sensor, which is denoted by the coordinate $y_0$. 

\begin{figure}[htbp]
  \centering
  \subfloat[]{\includegraphics[width=0.49\textwidth]{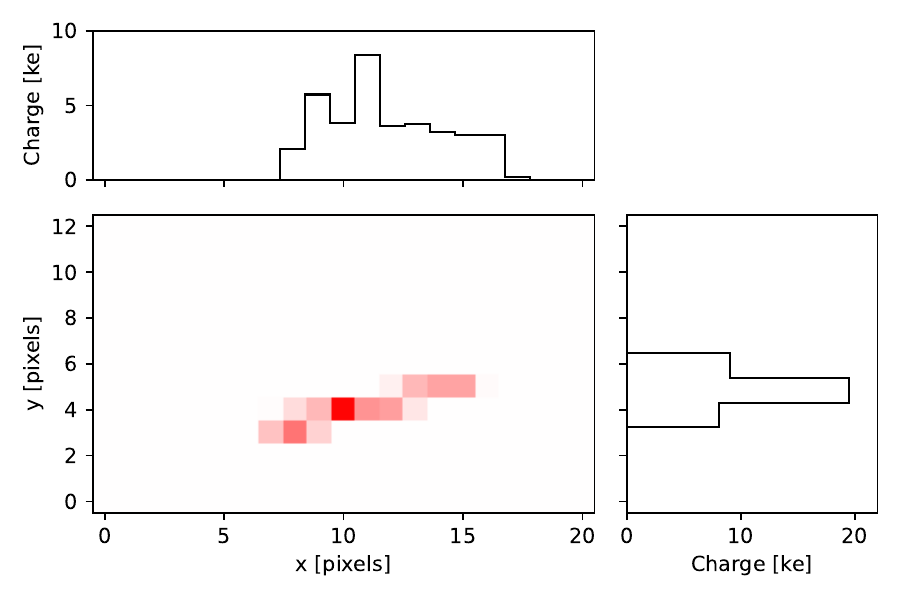}}
  \subfloat[]{\includegraphics[width=0.49\textwidth]{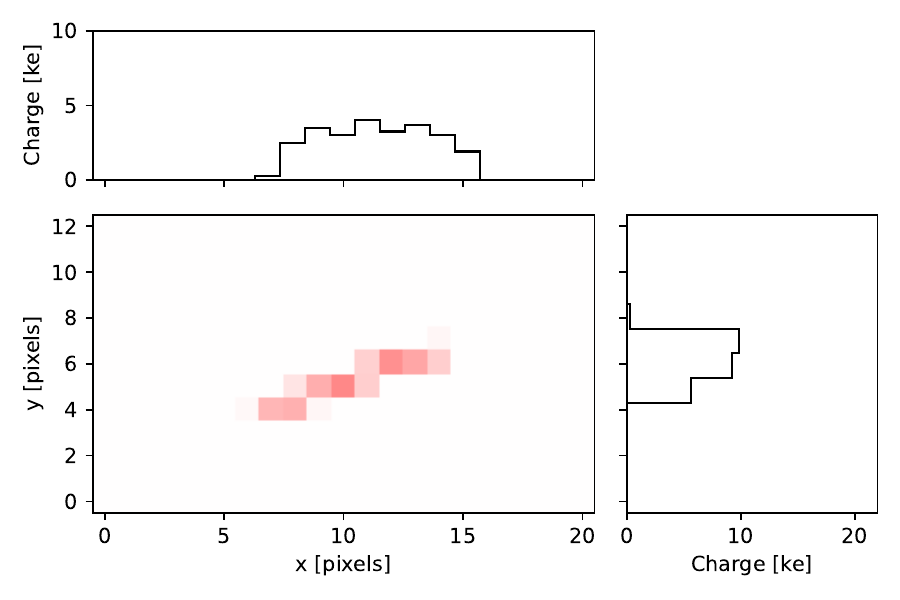}}
   \caption{Two example charge clusters and the corresponding $x$- and $y$-profile projections.  The color scale (common between the panels) represents the collected charge. Both clusters have $y_0=2.3$ mm, but different $p_T$: (a) $p_T=1.9$ GeV, (b) $p_T = 135$ MeV.}
  \label{profiles}
\end{figure}

For a given cluster, the sum over pixel columns projects the cluster shape onto the $x$-axis: this distribution is referred to as the \textit{x-profile}. The sum over pixel rows, \textit{y-profile}, which projects the cluster shape onto the $y$-axis, is sensitive to incident angle $\beta$ and therefore to the particle's $p_T$.  Two example clusters are shown in Figure~\ref{profiles} with the corresponding $x$- and $y$-profile projections.

\begin{figure}[htbp]
  \centering
  \subfloat[]{\includegraphics[width=0.33\textwidth]{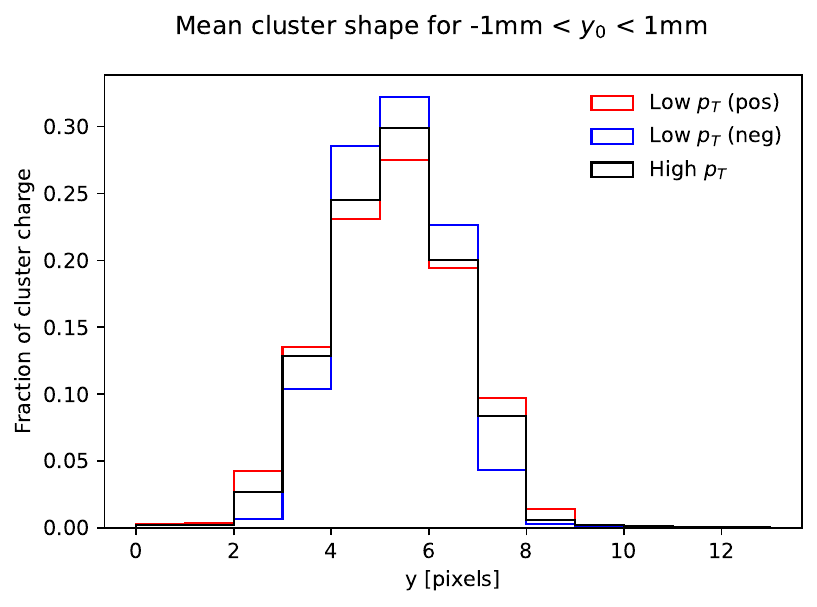}}
  \subfloat[]{\includegraphics[width=0.33\textwidth]{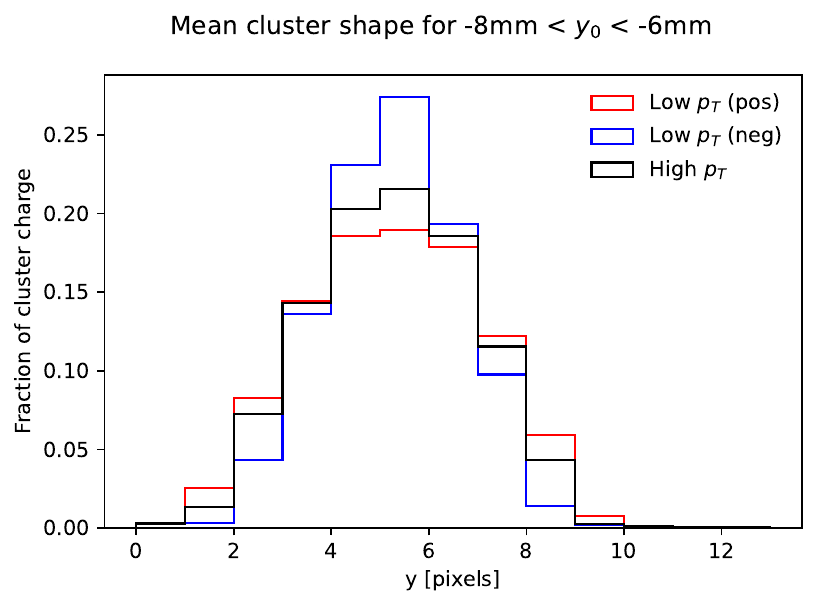}}
  \subfloat[]{\includegraphics[width=0.33\textwidth]{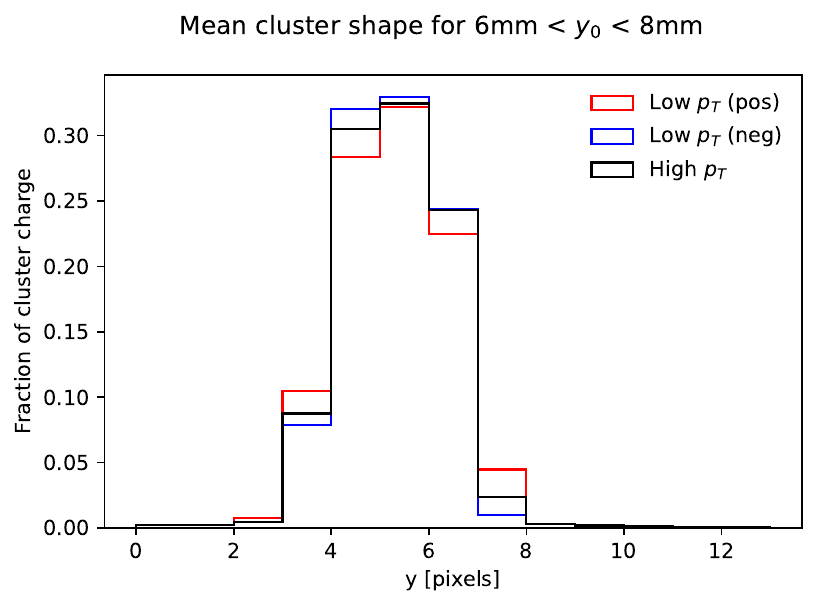}}
   \caption{Distribution of $y$-profile at different $y_0$: (a) $-1 < y_0 < 1$ mm, (b) $-8 < y_0 < -6$ mm, and (c) $6 < y_0 < 8$ mm. Shown separately for positively charged low $p_T$ particles (red), negatively charged low $p_T$ particles (blue) and high $p_T$ particles of both signs (black).}
  \label{yprofile}
\end{figure}

The mean $y$-profile cluster charge distributions for particles impinging near the center of a sensor ($-1 < y_0 < 1 $ mm ) are shown in Figure \ref{yprofile}a for three populations of clusters. Clusters created by high $p_T$ particles ($p_T > 2$ GeV) are represented by the black distribution, while clusters created by low $p_T$ particles ($p_T < 200$ MeV) are represented by red and blue for positively and negatively charged particles, respectively. Due to the deflection of charge carriers by the magnetic field (Lorentz drift), the cluster shape is not symmetric in $\beta$. For a flat detector module that measures 16 mm in $y$, particles of similar momentum leave markedly different cluster shapes at different $y_0$ positions on the module. Figures \ref{yprofile}b-c show the average cluster shapes at the extreme edges of the module. 

\begin{figure}[htbp!]
  \centering
  \includegraphics[width=0.7\textwidth]{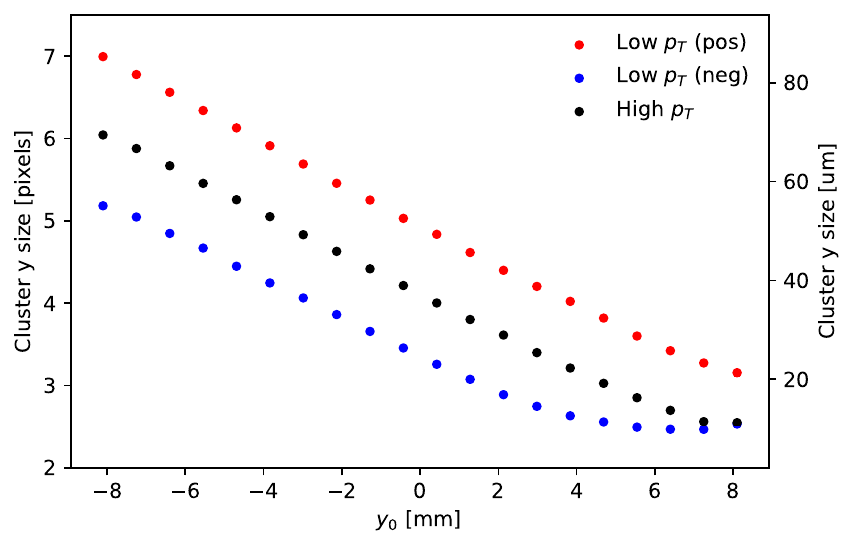}
   \caption{Cluster $y$-size vs. $y_0$ for different values of particle charge and $p_T$. The decrease in cluster size from the left to right side of the sensor plane is due to Lorentz drift.}
  \label{sizeVsYlocal}
\end{figure}

The cluster $y$-size is defined as the number of pixel rows in which non-zero net charge has been deposited after 4 nanoseconds. The dependence of the cluster $y$-size on both the charged particle $p_T$ and $y_0$ is shown in Figure \ref{sizeVsYlocal}. The decrease in $y$-size from the left to right edge of the sensor is due to Lorentz drift. In order to study the potential gain from timing information, the simulation also records the induced and collected charge in each pixel at 200 picosecond time intervals. 

\subsection{Untracked data}
\label{sec:untracked}

The simulated dataset described above is derived from clusters in the CMS detector that are combined with signatures in other detector layers to form particle tracks. In an example 2022 CMS data taking run, only about 40\% of clusters in the innermost layer of the CMS pixel detector are associated with tracks in this way. The remaining 60\% of clusters, designated as \textit{untracked}, can result from a variety of sources, including very low $p_T$ particles, radiation related backgrounds, or detector effects. 

Such untracked clusters are difficult to model, but critical to include in an estimate data reduction. A dedicated dataset has therefore been generated for this purpose based on untracked clusters in CMS data. The untracked clusters from CMS data are used as a proxy for backgrounds that should be rejected at a general collider experiment and are not being used to tune a specific pixel upgrade project. For each untracked cluster measured in CMS, we assume that it was created by a charged pion with energy 1 GeV traversing the detector.  The incidence angles $\alpha$ and $\beta$ of the particle are approximated based on the cluster size:
\begin{align}
\text{size}_x = \frac{t}{\Delta_x}|\cot\alpha| + 1 \\
\text{size}_y = \frac{t}{\Delta_y}|\cot\beta + \frac{y_0}{R}| + 1
\end{align}
Here, $t$ represents the sensor thickness and $\Delta_x$ and $\Delta_y$ the pixel pitch in the $x$ and $y$ directions. The degeneracy in the incidence angle due to the absolute value is broken by using measured correlations between $\beta$ and $y_0$ and $\alpha$ and $z$ to assign values to each cluster.  The $(x,y)$ position where the particle traverses the sensor mid-plane is assumed to be uniform across the central $3\times3$ pixel array.  The response of our futuristic pixel detector to charged pions with the corresponding properties is then simulated. 

The resulting description of untracked clusters is approximate, and a more precise description can be derived in the future using data from beam tests. 

\FloatBarrier

\section{Pixel cluster filtering benchmark and algorithm design}
\label{sec:algo}

The vast majority of particles produced at the LHC correspond to low-energy hadronic activity that is not pertinent to the studies of high-energy collisions. This is exacerbated by the increased rate of pileup at the HL-LHC compared to previous LHC runs. Rejection of clusters consistent with low $p_T$ particles at the data source, within the pixelated area of a readout chip, would save bandwidth by reducing the amount of data that must be transferred off-chip. 

This study focuses on filtering out tracks created by low momentum particles, and therefore uses only the $y$-profile information which is most relevant for track curvature in the magnetic field.  In future studies, we will look to exploit more of the cluster shape information for further data reduction.

\subsection{Single cluster $p_T$ filtering algorithm}
\label{sec:fullprecision}

A neural network classifier is designed to identify clusters initiated by high $p_T$ charged particles. The high $p_T$ signal class contains tracks with $p_T>200$ MeV. Because oppositely charged particle tracks curve in opposite directions in the magnetic field, the resulting clusters are very different in shape, as shown in Figure \ref{yprofile}. Therefore two background classes are defined, corresponding to positively and negatively charged particles with $p_T<200$\,MeV.  The choice of $p_T$ threshold in the training class definition affects the physics performance and could be adjusted depending on the physics goals. In this study, we find that a 200 MeV threshold in the training data results in a flat signal efficiency for track $p_T>2$\,GeV, which we consider to be the useful range for physics analysis. 

The simulated dataset of 800,000 clusters is used for training (no untracked data). The simulated dataset is split into a training set (80\%) and a test set (20\%) to be used for evaluation of the algorithm's performance. While the true $p_T$ distribution is shown in Fig.~\ref{pT}, the training sample provided to the network contains the same number of clusters for each $p_T$ class. 

All models were implemented in TensorFlow~(\texttt{v2.10.0})~\cite{tensorflow2015-whitepaper} using the Keras API~\cite{chollet2015keras}. Neural network trainings were run for 200 epochs where early stopping was used if the loss function showed no improvement after 20 epochs. A batch size of 1024 was used in all models. The Adam optimizer \cite{adamoptimizer} with a learning rate of 0.001 was used in conjunction with the Keras Sparse Categorical Cross entropy loss function in all models. The models are trained with three output categories: positively charged and $p_T<200$\,MeV; negatively charged and $p_T<200$ \,MeV; and $p_T>200$\,MeV, both positively and negatively charged.  A softmax activation was used in all models to generate classification probabilities between 0 and 1, and each cluster is assigned the classification label corresponding to the highest probability.

The charge cluster shape along the axis parallel to the magnetic field direction (sensor $x$) is assumed to be largely uncorrelated with track $p_T$. The projection of the cluster shape onto the sensor $y$-axis ($y$-profile) is therefore used as the training input to the classifier. Three training setups are developed corresponding to input features of different complexity in order to demonstrate how additional information improves $p_T$ discrimination:
\begin{itemize}
    \item Cluster $y$-size: number of pixel rows with nonzero charge deposited after 4\,nanoseconds
    \item Cluster $y$-profile: the amount of charge collected in each row of pixels after 4\,nanoseconds
    \item Cluster $y$-profile with timing: the amount of charge collected in each row of pixels evaluated at eight intervals of 200\,picoseconds
\end{itemize}
In each case, the neural network can be trained using the cluster position on a flat module ($y_0$) as an additional input feature. Alternatively, separate networks can be trained for each region in $y_0$ (see Section \ref{design-space}). Each model is summarized below. 

\paragraph{Model 1: cluster $y$-size.}Two input features are used in Model 1: the cluster $y$-size and position $y_0$. The model consists of one dense layer with 128 neurons and 771 parameters.  This model provides a test of performance with minimal information provided to the neural network. 

\paragraph{Model 2: cluster $y$-profile.}The cluster $y$-profile model has fourteen input features: cluster $y$-profile (thirteen features corresponding to thirteen pixel rows) and the $y_0$ position (1 feature). This model consists of one dense layer with 128 neurons and 2307 parameters.  

\paragraph{Model 3: cluster $y$-profile with timing information.}The third and most complex model takes as input the cluster $y$-profile distribution at eight time slices ($13 \times 8$ features) and the $y_0$ position (1 feature). The first eight time slices contain the most useful information, as most charge deposition occurs at the beginning of the cluster time evolution. This model uses a convolutional neural network (CNN) to pass a time-lapse picture of the cluster charge to the network. The cluster $y$-profile inputs weree passed through two two-dimensional convolutional layers (Conv2D), with 16 and 64 filters, respectively, using ReLU activations to introduce non-linearity \cite{agarap2018deep}. The shape of the kernels was $3 \times 3$, and strides was $1 \times 1$. The output of the Conv2D layers was flattened and concatenated with the $y_0$ input. This was then passed through a dense layer with 32 neurons, and using dropout of 0.1. The final model contains 83,331 parameters.  

\begin{figure}[htbp]
  \centering
  \includegraphics[width=0.5\textwidth]{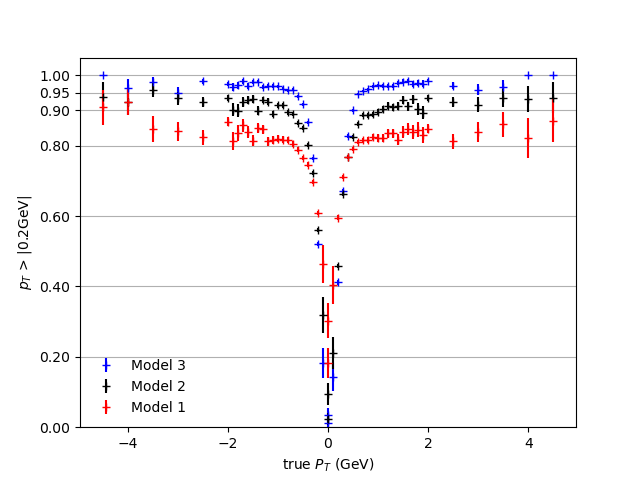}
   \caption{Classifier acceptance as a function of $p_T$ for three models with different input features. Positive and negative values of $p_T$ represent the performance on clusters initiated by particles of positive and negative charge, respectively.}
  \label{efficiency}
\end{figure}

The classifier acceptance is defined as the fraction of clusters that the network selects as $p_T>200$ MeV. 
A comparison of the three models in terms of the acceptance is shown as a function of the true $p_T$ in Figure \ref{efficiency}. 
Table \ref{reduction} compares two figures of merit for each model:
\begin{itemize}
\item signal efficiency:  the fraction of clusters with $p_T>2$ GeV that are classified as high $p_T$     
\item background rejection: the fraction of clusters with $p_T<2$ GeV that are classified as low $p_T$ 
\end{itemize}

\begin{table}[!htp]
\begin{center}
\begin{tabular}{llll}
\textbf{Model} & \textbf{Sig. efficiency} & \textbf{Bkg. rejection} \\ \hline 
Model 1 & 84.8 \% & 26.6 \%   \\ 
Model 2 & 93.3 \% & 25.1 \%  \\ 
Model 3 & 97.6 \% & 21.7 \%  \\ 
\hline
\end{tabular}
\caption{Comparison of model performance in terms of signal efficiency and background rejection.}
\label{reduction}
\end{center}
\end{table}

Model 1 ($y$-size and $y_0$) has the simplest architecture and achieves the highest data reduction rate. However, the information contained in only two features is insufficient for achieving a high signal efficiency, and this model selects less than 85\% of tracks with $p_T>2$ GeV.  Model 2 (cluster $y$-profile) achieves an accuracy of $93.3\%$ for tracks with $p_T>2$ GeV, and remains sufficiently compact for implementation on-ASIC. The inclusion of timing information in Model 3 achieves an additional 4\% gain in the signal efficiency and is the most accurate overall. However, the extraction of time-sliced charge information presents challenges to the chip architecture that merit further study but remain outside the scope of this work. \emph{Model 2 is therefore chosen as the baseline model for hardware implementation.}

\subsection{Estimate of overall reduction in bandwidth}
\label{sec:data_reduction}

The full dataset that will be seen by our detector is a combination of the simulated dataset described in Section \ref{sec:tracked}, the untracked dataset described in Section \ref{sec:untracked}, and the single pixel clusters that are not associated with tracks. Due to the small pixel size, all simulated clusters resulting from the traversal of a charged particle consist of at least two pixels. We can therefore assume that single pixel clusters will be rejected out of hand. 

The first column of Table \ref{breakdown} shows the expected contribution from each component sample based on a 2022 CMS data run. The second column shows the fraction of clusters in each sample that are rejected by the baseline model (Model 2 in Section \ref{sec:fullprecision}). The fraction of simulated clusters rejected is corrected to account for the CMS tracking efficiency. 

\begin{table}[!htp]
\begin{center}
\begin{tabular}{lll}
\textbf{} & \textbf{Fraction of dataset} & \textbf{Rejection rate} \\ \hline
Simulated tracks & 40\% & 36.3\% \\
Multi-pixel untracked & 55\% & 61.9\% \\
Single pixels & 5\% & 100\% \\
\hline
\end{tabular}
\caption{Breakdown of the total dataset seen by the detector and the rejection rate achieved on each subsample. \label{breakdown}}
\end{center}
\end{table}

Because the classifier training is agnostic to the multi-pixel untracked dataset, we consider the 61.9\% rejection achieved on this sample to be a conservative lower bound. Using this lower bound, the expected fraction of clusters rejected by the classifier is 53.5\%. Assuming that 100\% rejection can be achieved on multi-pixel untracked clusters gives an upper bound on the fraction of clusters rejected of 74.5\%. 

Charge clusters vary in size, and the size of the data read out per cluster is proportional to the number of pixels contained in the cluster. The overall reduction in bandwidth is therefore calculated by weighting each cluster by the number of pixels it contains. The result is a bandwidth savings of 54.4-75.4\%. 
\subsection{Model quantization}
\label{sec:model_quantization}



The training inputs of the models discussed thus far are the full-precision $y$-profile distribution (the exact number of electrons simulated in each pixel) and a single-precision floating point value for the $y_0$ position. However, in order for the $p_T$ filtering algorithm to satisfy the power and area requirements for on-ASIC implementation, the charge collected per pixel must be described by the output of an analog-to-digital converter (ADC). The baseline quantized model uses a 2-bit quantization of $y$-profile with each output corresponding to a range of collected charge, as summarized in Table \ref{quantized-inputs}. The optimization of this input quantization and the size of the ADC is discussed later in this Section. The $y_0$ coordinate is reduced to a 6-bit input by dividing the range of possible $y_0$ values into 64 equal bins of 250\,$\mu m$ width. 

\begin{table}[!htp]
\begin{center}
\begin{tabular}{ll}
\textbf{ADC output} & \textbf{Charge interval [$e^-$]} \\ \hline
00 & $<400$ \\
01 & $400-1600$ \\
10 & $1600-2400$ \\
11 & $>2400$ \\
\hline
\end{tabular}
\caption{Mapping between 2-bit ADC output and collected charge.}
\label{quantized-inputs}
\end{center}
\end{table}

\begin{figure}[htbp]
  \centering
  \includegraphics[width=0.5\textwidth]{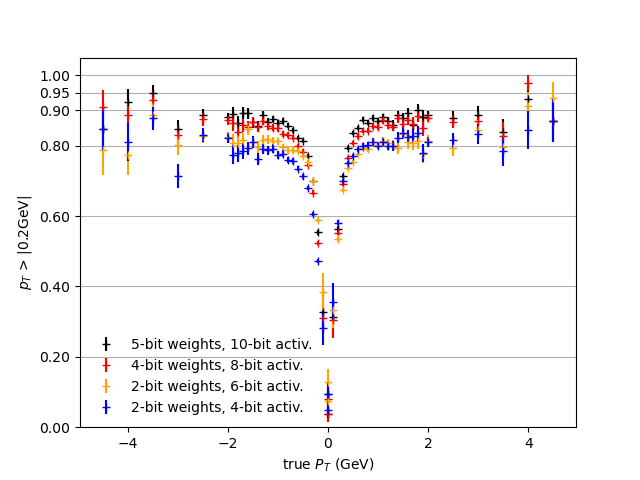}
  \caption{Classifier acceptance vs. $p_T$ for Model 2, with each curve corresponding to a different choice of quantization hyper-parameters. Positive and negative values of $p_T$ represent the performance on clusters initiated by particles of positive and negative charge, respectively.}
  \label{quantized-efficiency}
\end{figure}

For implementation on-ASIC, the neural network weights must also be quantized to a precision of a few bits without significant loss in performance. The QKeras library \cite{Coelho_2021,https://doi.org/10.48550/arxiv.2102.04270} is used to perform quantization-aware training, enabling an early evaluation of the impact of low bit precision on the model's performance. This allows an initial assessment of the trade-offs between accuracy and resource utilization before finalizing a design for the ASIC implementation (see Sec \ref{design-space}).  The quantized baseline model consists of a quantized dense layer and a quantized representation of a ReLU activation function. Additionally, a Batch Normalization (BN) layer that provides a regularization effect during training, akin to dropout, was incorporated to prevent overfitting. A softmax activation generates classification probabilities between 0 and 1, and each cluster is assigned the classification label corresponding to the highest probability. The final quantized model has 2819 parameters. 
The number of bits used for network weights and activation layers were treated as hyper-parameters of the model, and  chosen to maximize the signal efficiency. Figure \ref{quantized-efficiency} compares the performance for selected combinations of the quantization hyper-parameters following a hyperparameter search.  

\begin{figure}[htbp]
  \centering
  \includegraphics[width=1.\textwidth]{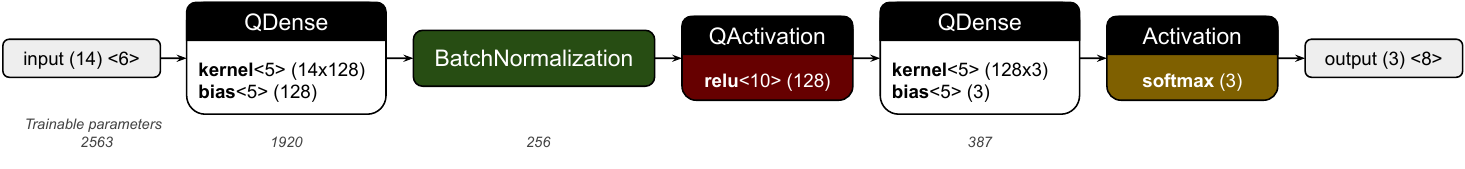}
  \caption{Baseline quantized model consisting of 14 inputs (13 bins of $y$-profile plus $y_0$), three outputs, and two hidden layers containing 128 and three neurons each. For each layer, the dimension is reported in round brackets and the bit-width of the fixed-point representation in angular brackets.} 
  \label{fig:model_dse_first}
\end{figure}

The model achieving best performance uses five-bit weights and ten-bit activations and is illustrated in Figure \ref{fig:model_dse_first}. However, the model using four-bit weights and eight-bit activations gives only slightly lower efficiency, and is ultimately selected for hardware implementation given resource constraints (see Section \ref{design-space}).

The full precision baseline model accurately classifies 93.3\% of tracks with $p_T>2$ GeV as high $p_T$. Quantizing the collected charge and $y_0$ inputs reduced this accuracy to 88.8\%. The quantization of the neural network weights and activations via QKeras led to a further drop in accuracy of about 1.5\%, suggesting that the quantization via QKeras was efficient. 
Table \ref{reduction-quantized} compares the performance of the full precision model, the model with quantized input features only, and the model with quantized input features and weights.


\begin{table}[!htp]
\begin{center}
\begin{tabular}{llll}
\textbf{Model}  & \textbf{Sig. efficiency} & \textbf{Bkg. rejection} \\ \hline 
Full precision &  93.3 \% & 25.1 \%  \\ 
Quantized inputs & 88.8 \%  & 25.8 \%  \\
Quantized weights \& inputs & 87.3 \% & 28.2  \%  \\
\hline
\end{tabular}
\caption{Comparison of the baseline model performance at different stages of the quantization in terms of signal efficiency and background rejection.}
\label{reduction-quantized}
\end{center}
\end{table}

\FloatBarrier
\subsubsection*{Noise threshold}

Because the $p_T$ filtering algorithm is sensitive to cluster shape, noise hits can be rejected based on both shape and magnitude of collected charge. This allows for the noise threshold (the boundary between ADC output 00 and 01) to be set lower than in present-day pixel detectors, where a typical threshold of $\sim1500$ electrons is selected to suppress noise. 

Table \ref{noise-threshold} shows the signal efficiency and background rejection for several example noise thresholds. Setting lower thresholds tends to improve performance as it allows more information to be passed to the neural network classifier. However, because the input $y$-profile distribution represents a sum over pixel columns, the classifier is not especially sensitive to this threshold. The 400 electron threshold was chosen for the baseline model as it provided good performance and could serve as a realistic threshold for hardware.
\begin{table}[!htp]
\begin{center}
\begin{tabular}{llll}
\textbf{Threshold [$e^-$]} &\textbf{Sig. efficiency} & \textbf{Bkg. rejection} \\ \hline 
400 &  87.7 \% &  27.0 \% \\ 
600 &  86.0 \% &  28.2 \% \\ 
800 &  86.0 \% &  27.1 \% \\ 
\hline
\end{tabular}
\caption{Signal efficiency and background rejection  for quantized Model 2 using different values of the noise threshold.}
\label{noise-threshold}
\end{center}
\end{table}

\subsubsection*{Number of bits in the ADC}
\begin{figure}[htbp]
  \centering
  \includegraphics[width=0.49\textwidth]{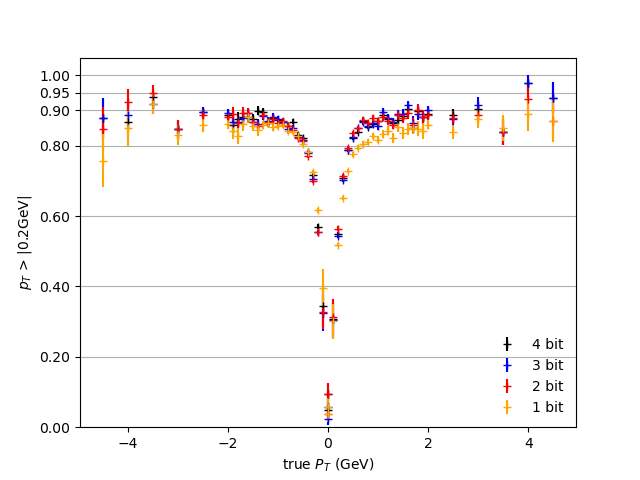}
   \caption{Classifier acceptance vs. $p_T$ for the quantized Model 2 using different numbers of bits in the ADC. Positive and negative values of $p_T$ represent the performance on clusters initiated by particles of positive and negative charge, respectively.}
  \label{adc-bits}
\end{figure}

The baseline quantized model uses a 2-bit ADC where the output corresponds to the charge intervals in Table \ref{quantized-inputs}.  The model was also trained with 1, 2, 3, and 4 bits in the ADC using multiple mappings of charge intervals to ADC value. Figure \ref{adc-bits} shows a comparison of the best performing quantized model for each number of bits in the ADC. Details of the alternative mappings are included in \ref{app:adc-bits}.

Because the 2-bit ADC provides many advantages for low-power implementation, with minimal loss of performance in the $p_T$ filter, this option is selected for the final chip.  

\FloatBarrier

\section{On-chip Implementation}
\label{sec:implementation}

Hybrid pixel detectors consist of silicon sensors bonded to a pixelated readout integrated circuit (ROIC). The ROIC is designed in a high performance 28\,nm CMOS process. It amplifies and digitizes the signal produced by charged particles traversing the detector and selectively transmits data off-chip for further analysis. In this section, we explore implementation of the baseline algorithm within the pixelated region of the ROIC to enable data-filtering at source, i.e. the rejection of data clusters from low momentum particles.  

Each ROIC pixel contains an analog frontend consisting of a charge sensitive preamplifier and three auto-zero comparators for a 2-bit flash-type ADC operating synchronously at 40 MHz.  This makes it possible for all hits to be accurately recorded, even when two hits occur in consecutive bunch crossings, without off-time registration of events. A more detailed description of the 2-bit ADC which provides the input to the algorithm is described in Ref.~\cite{ISCAS2023}.


\subsection{ROIC mapping to sensor geometry} 

\begin{figure}[htbp]
  \centering
  \includegraphics[width=0.6\textwidth]{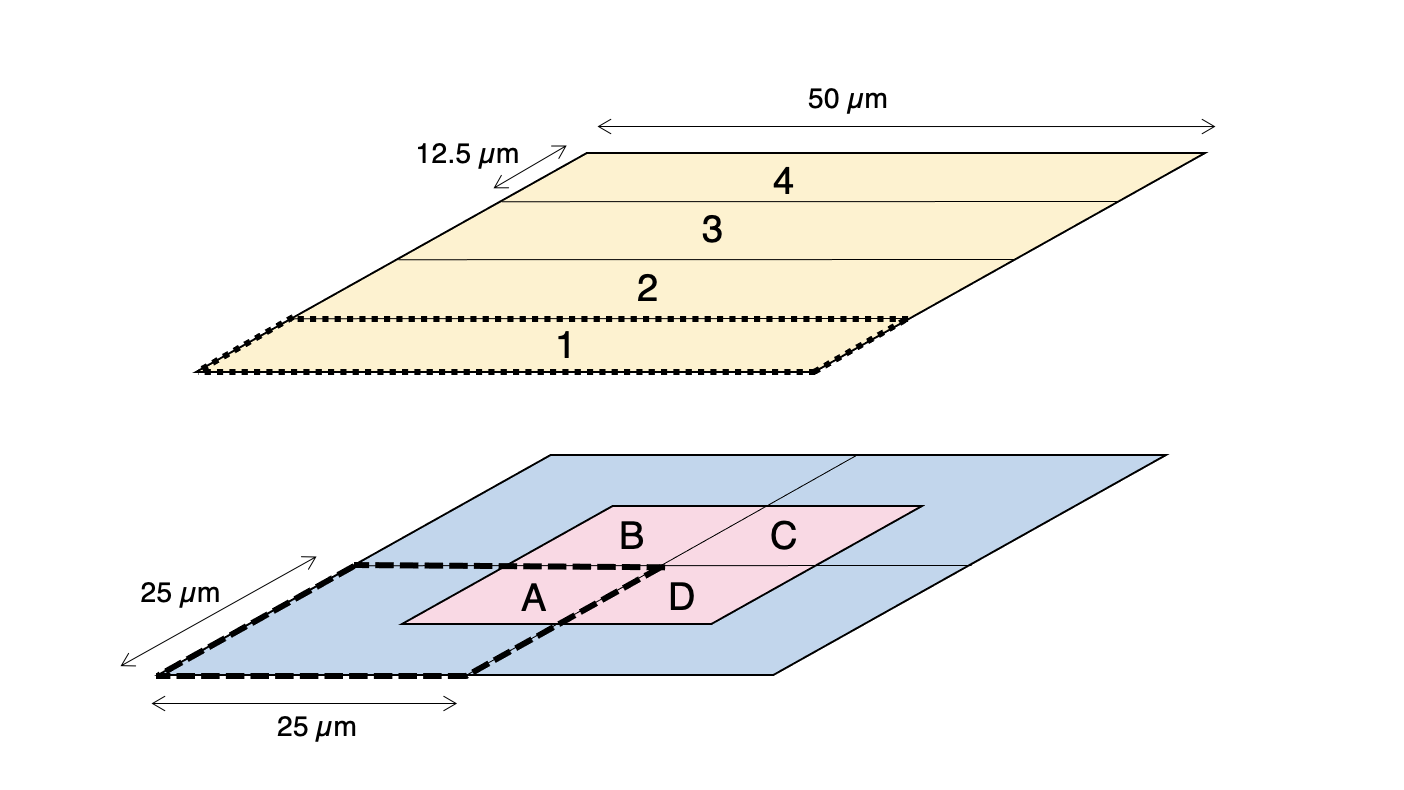}
   \caption{A diagram of the mapping of sensor pixels (yellow) to ROIC pixels (analog portion in pink, digital in blue).}
  \label{roic}
\end{figure}

A $2\times2$ array of ROIC pixels (corresponding to 50$\times$50\,$\mu m$) is laid out so that the four analog frontends create an island, which is surrounded by the digital logic of the ADC and the filtering algorithm. This geometry is shown in Figure \ref{roic}, where the pink (blue) region represents the analog (digital) region of the ROIC pixel. This layout, which is a common technique used in pixel detectors, maximizes efficiency in terms of area ~\cite{Sproket2023,RD53Bspec}.

Each $2\times2$ array of ROIC pixels corresponds to a $1\times4$ array of the sensor pixels described in Section \ref{sec:dataset}. The sensor pixels are shown in yellow in Figure \ref{roic}, and the ROIC-to-sensor pixel mapping links ROIC pixels A, D, B, C to sensor pixels 1, 2, 3, 4 respectively. 

The clusters used for algorithm development discussed in previous sections are simulated in a 13\,$\times$\,21 array of sensor pixels. However, to simplify the digital implementation, we assume a 16\,$\times$\,16 array of sensor pixels for the ROIC. The size of each sensor pixel remains unchanged with a dimension of 50\,$\times$\,12.5\,$\mu m$. A \textit{super-pixel} of 16\,$\times$\,16  sensor pixels will be bonded and remapped to a matrix of 4\,$\times$\,32 frontent electronics pixels as shown in Figure~\ref{roic}. This does not change the performance of the algorithm except in the case of very wide clusters (in the bending plane). However, it is still straightforward to validate the algorithm digital implementation. 

\subsection{Design space optimization}
\label{design-space}

Following the selection of a baseline quantized model in Sec.~\ref{sec:model_quantization}, an additional architechtural exploration was performed that fully accounts for hardware constraints. During this stage, five major design decisions were made in order to compress the model without significant loss of performance: 
\begin{enumerate}
    \item the BN layer is folded into the Dense layer
    \item the number of neurons is reduced
    \item the number of bits in the network weights and activations is reduced
    \item the final activation is changed from Softmax to Argmax. 
    \item $y_0$ is not directly used as an input to the model. Instead we employ a region-specific implementation.
    \end{enumerate}
Each of these design choices is described in detail below. 

First, 
we folded the BN layer into the Dense layer. This optimization technique combines the BN parameters with the Dense layer's weights and biases, effectively reducing computational costs during inference while maintaining the model's performance~\cite{jacob2018quantization}. The resulting model has 2,307 trainable parameters, which is a decrease from 2,563 trainable parameters in the original model.

Second, we examined the trade-off between accuracy and area by considering the reduction in the number of neurons and bit precision. Reducing the number of neurons in a neural network can lead to decreased accuracy due to diminished model capacity, resulting in the loss of vital information, underfitting, and limited expressiveness. However, in terms of hardware implementation, fewer neurons create a simpler data path, thereby reducing the required area. In our exploration, we sought to balance network complexity and hardware area while striving to maintain the accuracy of the original network outlined in Sec.~\ref{sec:model_quantization}. The optimal model was found to reduce the number of neurons of the first dense layer from 128 to 58. 

Third, we reduced the number of bits in the weights and activations by 2 bits each. Design decisions ii (reduction in number of neurons) and iii (reduction of the bit-width of the fixed point representation) produced a hardware implementation with a third of the area originally necessary with a drop in signal efficiency of only 3.5\%. 

To further reduce the computational complexity of the model without sacrificing the overall predictive capability, we utilized an Argmax function as the final layer (decision iv) instead of the conventional Softmax activation function. By directly identifying the class with the highest score using the Argmax function, we bypass the need to calculate the probability distribution across all classes, as required by the Softmax function.

\begin{figure}[htbp]
  \centering
  \includegraphics[width=1.\textwidth]{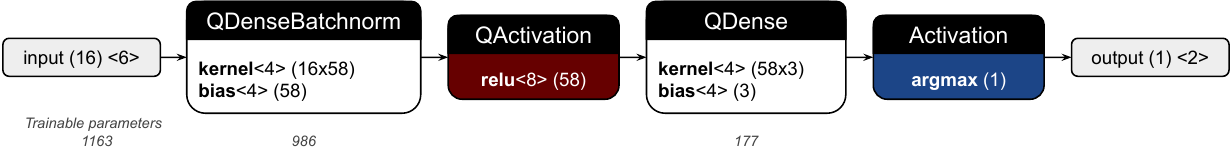}
  \caption{Final model architecture with reduced bit-precision and trainable parameters.  For each layer, we report the dimensions in round brackets and the bit-width of the fixed-point representation in angular brackets.} 
  \label{fig:model_dse}
\end{figure}

Finally, we opted to remove the $y_0$ coordinate from the model entirely. Rather than directly providing the network with the value of this coordinate, we train the baseline network many times on clusters in restricted ranges of $y_0$: this is referred to as the \textit{region specific implementation}.  The network architecture in each $y_0$ region is identical, but the values of the reprogrammable network weights can be tuned based on the ASIC's $y_0$ position. The input $y$-profile distribution is then expanded from 13 to 16 bins by padding with zeros, so that the pixel array can be comprised of a round number of the ROIC units shown in Figure \ref{roic}. 
Fig.~\ref{fig:model_dse} depicts the final model architecture, with all updates to the design incorporated. 

\begin{figure}[htbp]
  \centering
  \includegraphics[width=0.6\textwidth]{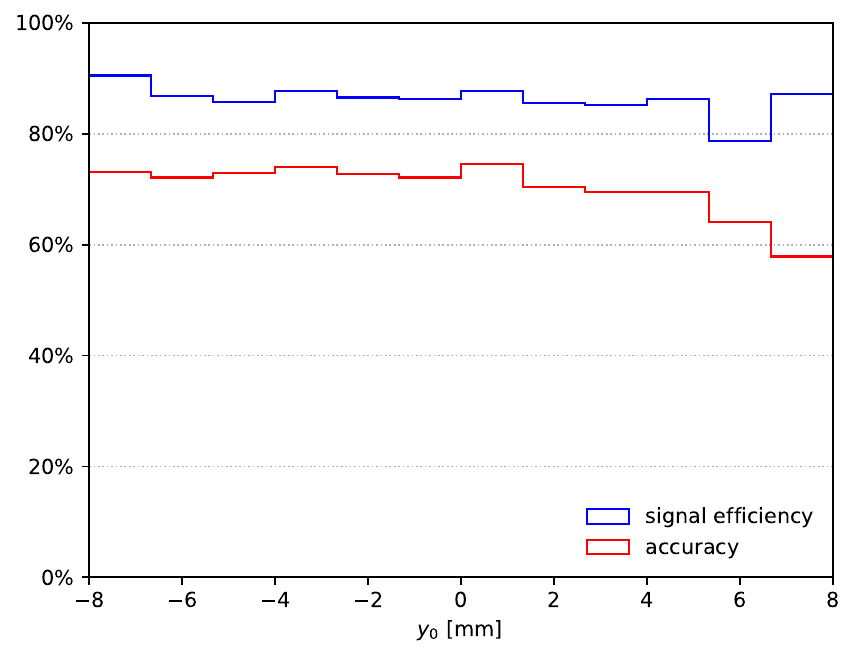}
  \caption{Signal efficiency and accuracy as a percentage in different $y_0$ regions for the final architecture to be implemented on-ASIC.}
  \label{fig:accuracy_per_ylocal}
\end{figure}

Figure \ref{fig:accuracy_per_ylocal} shows the signal efficiency and accuracy of our final architecture trained in different ranges of $y_0$. A decrease in accuracy is observed at the edge of the sensor ($y_0>6$mm), where Lorentz drift effects are the greatest and the cluster size decreases to an average of only two pixels, making the shapes of low and high $p_T$ tracks nearly indistinguishable (recall Fig.~\ref{sizeVsYlocal}).

The final optimized model has 1,163 trainable parameters, which is a reduction of 55\% with respect to the original model. 

\subsection{Design of algorithm implementation with \texttt{hls4ml}}
\label{design-space}

To translate the algorithm from a quantized graph representation into an optimal hardware implementation, we use the \texttt{hls4ml} workflow, an open-source Python framework for co-design and translating machine learning algorithms into hardware implementations~\cite{vloncar_2021_5680908,Duarte:2018ite}. The \texttt{hls4ml} workflow begins with a trained model from a conventional machine learning framework such as TensorFlow or PyTorch, or a quantized model from QKeras. Designers can use  \texttt{hls4ml} to optimize not only numerical precision, but also the hardware implementation's parallelism, striking a balance between area, performance, and power consumption based on the system constraints~\cite{fahim2021hls4ml}. Subsequently, \texttt{hls4ml} translates the model into C++ code for Siemens Catapult HLS~\cite{catapult-hls}. The HLS tool generates a hardware description at the register-transfer level (RTL) for the traditional ASIC flow. We have opted to fully parallelize the hardware logic, resulting in a combinational implementation of our models from HLS. This choice is driven by the desire to minimize the latency of the neural network. We integrated the resulting RTL design into the system alongside registers and data movers.

QKeras allows the designers to define and train models with specified bit precision for weights and activation functions, but lacks the capability to configure the bit precision of internal multiply-accumulators for different network layers, as it relies on floating-point representations. When converting the model to a C++ specification for HLS using \texttt{hls4ml}, the designers automatically import the fixed-point precision for weights and activations from QKeras. Additionally, they must manually specify the accumulator bit precision for each layer to prevent overflow and loss of precision during fixed-point arithmetic operations.
A careful choice of bit precision is crucial for multiply-accumulators, and \texttt{hls4ml} offers two potential solutions: a dynamic and a static approach. The dynamic approach requires monitoring the bit precision during simulation and adjusting the fixed-point format to avoid overflow and loss of precision. If the result exceeds the representable range or loses significant precision, an overflow or loss of precision has occurred. Designers must then adjust the fixed-point format by increasing the number of integer bits or fractional bits and repeat the simulation. The static approach determines the required bit precision without simulation, based on the given fixed-point precision for weights and activations. In \texttt{hls4ml}, an optimizer pass calculates the appropriate number of integer bits $n$ and fractional bits $m$ for a multiply-accumulator $Qn.m$ based on the bit precision of the two operands $Qn_1.m_1$ and $Qn_2.m_2$. For addition operations, it allocates $n = \rm{max}(n_1, n_2) + 1$ integer bits and $m = \rm{max}(m_1, m_2)$ fractional bits; for multiplications, it allocates $n = n_1 + n_2$ integer bits and $m = m_1 + m_2$ fractional bits. It is worth noting that a simulation-based approach tends to be dependent on the specific training results used for the simulation, and employing different weights from a subsequent training campaign may still result in overflow or loss of precision. In contrast, a static approach is more conservative and may require additional area to accommodate worst-case scenarios, but it is guaranteed to be error-free. In our exploration, we adopted the described static approach.

\begin{figure}[htbp]
  \centering
  \includegraphics[width=0.9\textwidth]{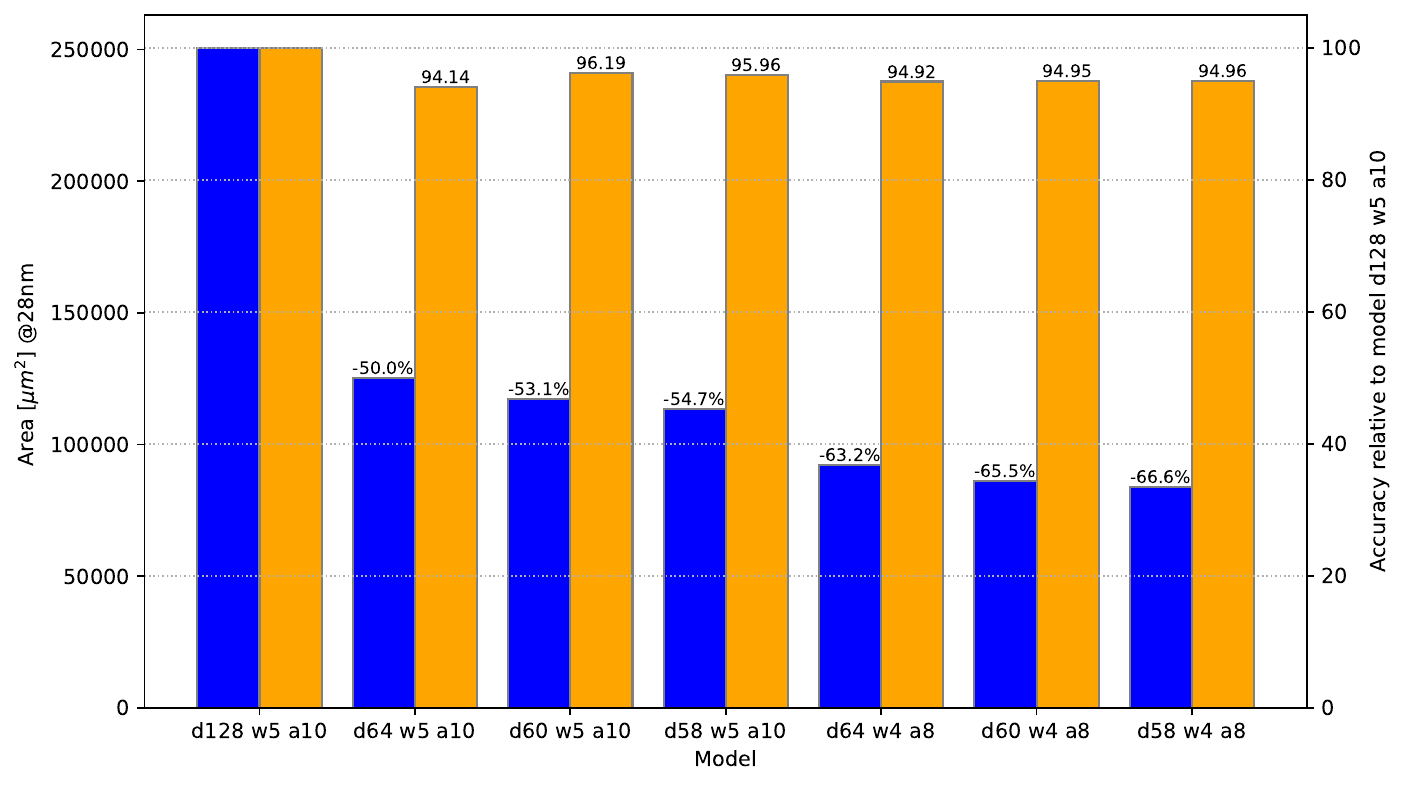}
   \caption{Comparison of the algorithm accuracy and area in 28nm CMOS for different model configurations. The model name corresponds to d\{number of neurons in the first hidden layer\} w\{weight bit-width\} a\{activation bit-width\}.}
  \label{fig:area}
\end{figure}

A summary of the hardware design space exploration is shown in Fig.~\ref{fig:area}.  Here we show the original 128-node hidden layer implementation and then the relative performance (accuracy) and digital area as we vary the size of the hidden dimension and the bit precision of the neural network computation.  We find that reducing precision and number of hidden nodes reduces the area by 67\% with a reduction of 5\% in accuracy.  We leave it to future work to continue optimization of the algorithm while this version of the algorithm is sufficient for digital implementation.   

\subsection{ROIC integrated sensing and edge computing}

Finally, the ML algorithm must be integrated into the array of ROIC pixels. The front-end of the ROIC must amplify and digitize the event's signal at 40 MSPS (Mega Samples Per Second) so the neural network can make an inference within a single bunch crossing. The architecture used is based on a synchronous flash ADC~\cite{ISCAS2023} since the conventional Time Over Threshold (TOT) architecture would require multiple clock cycles to digitize an event.

\begin{figure}[htbp]
  \centering
  \includegraphics[width=0.9\textwidth]{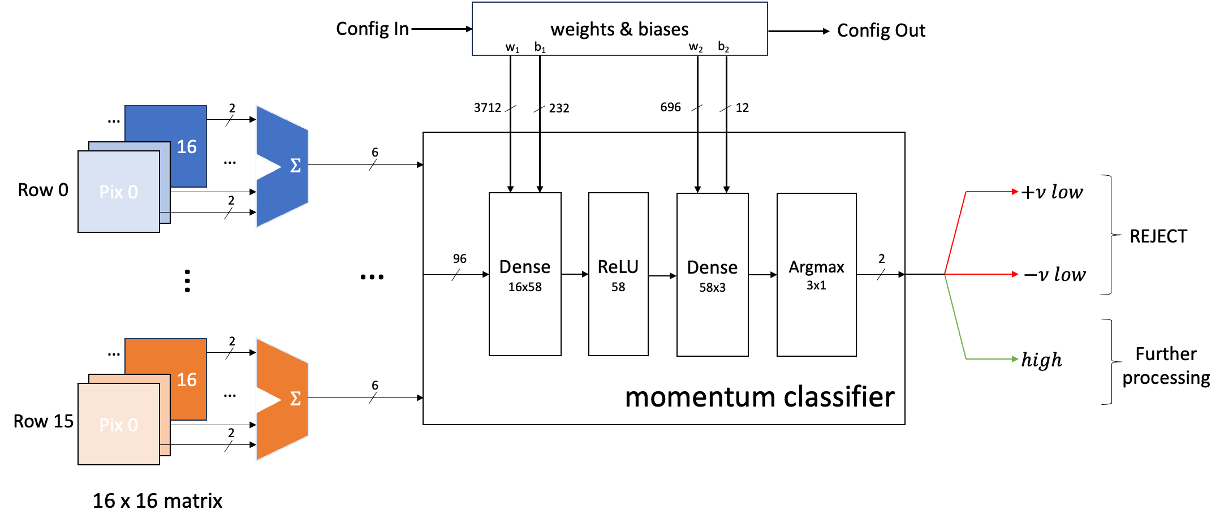}
   \caption{Data flow through the digital implementation of the algorithm from the summed ADC bits (on the left) through the neural network and the final classification layer. At the top of the diagram we illustrate the reconfigurability of the weights and biases in the algorithm stored in memory.}
  \label{fig:flow}
\end{figure}

The input to the algorithm, the $y$-profile distribution, is generated by summing the two-bit ADC outputs across all columns.  The algorithm accepts 16 inputs of size six bits each, corresponding to the $y$-profile distribution (13 inputs) padded with zeros for the implementation (see Section~\ref{design-space}). Ultimately, the two-bit output of the algorithm classifies the clusters as a negatively charged low momentum track, a positively charged low momentum track, or a high momentum track. The sum over the ADCs and their subsequent input into the NN algorithm is depicted in Fig.~\ref{fig:flow}.  At the top of the figure, we also illustrate that the algorithm is reconfigurable by introducing new weights and biases into a fixed architecture.  This allows the algorithm to be adapted to different regions of the detector and changing detector conditions (due to radiation damage, for example).

\begin{figure}[htbp]
  \centering
  \includegraphics[width=0.9\textwidth]{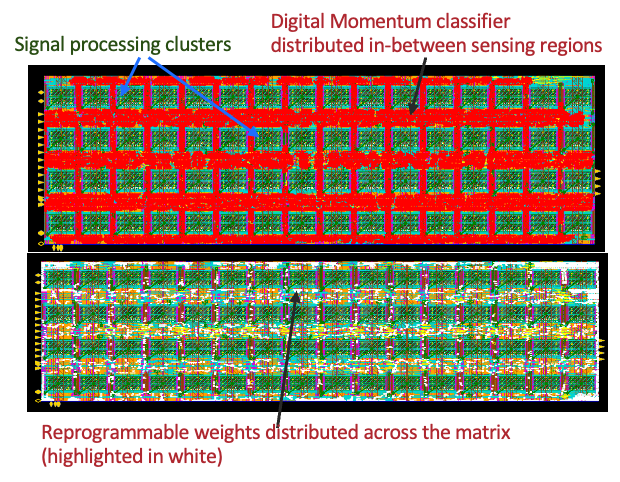}
   \caption{Physical layout of the 16\,$\times$\,16 pixel array digital implementation. The top view shows the analog islands in green and the digital implementation of the neural network in red. The bottom view shows different digital circuit elements highlighting the reprogrammable weight storage in white.}
  \label{fig:cloud}
\end{figure}

The physical layout of the super-pixel is shown in Figure~\ref{fig:cloud}. The size of the digitally implemented super-pixel is 889\,$\mu m$ $\times$ 222\,$\mu m$.
The green areas correspond to the analog circuit islands, while the red contains the digital logic.
Given our design, the registers require a one-time setup and leakage at low temperatures is deemed insignificant, thus the combinatorial logic is projected to account for the majority of the power utilization.
Based on the technology power models, the anticipated power consumption of the digital logic is roughly 300\,$\mu W$, measured when toggling 50\% of the inputs every clock cycle (25 nanoseconds).


%
%
%



\section{Conclusions and Outlook}
\label{sec:outlook}

High granularity silicon pixel sensors are at the heart of energy frontier particle physics collider experiments and provide the highest spatial granularity measurements for charged particles.  At an LHC collision rate of 40\,MHz, these detectors create massive amounts of data.  Our goal in this proof-of-concept study is to explore the potential for on-sensor data filtering to reduce the data rate from pixel sensors so that they could be read out at 40\,MHz and their data could be used in online trigger systems.  
To accomplish this task, we explore the lossy data compression task of filtering out clusters in the pixel detector originating from low momentum charged particle tracks by deploying on-sensor ML techniques.  This study sets a first baseline for what is possible using current technologies, and follow up studies will explore emerging technologies which could improve computational efficiency, physics performance, and additional data reduction techniques.  

This study makes a number of novel contributions to this task: 

\begin{itemize}
    \item We introduce a new public dataset which can be used for the benchmark task of pixel on-sensor data reduction.  It includes pixel sensor charge deposition information for realistic track distributions as well as  information about the charge deposition time structure which can potentially be used in emerging computing architectures.  
    \item We develop and compare a number of algorithmic approaches from simple cluster size information to cluster distributions to time-evolved cluster distributions and present their performance for the cluster filtering task.  This shows that for the simple cluster distibution inputs, we can filter out, conservatively, 54\% of the pixel sensor data while maintaining a $>$\,90\% efficiency for clusters from tracks with $p_T>$\,2\,GeV. We then optimize the algorithm for circuit implementation by exploring quantization of both the charge deposition inputs and the neural network parameters.
    \item As a first design for implementation on chip of these algorithms, we synthesize and integrate the algorithm which takes as input the cluster distribution ($y$-profile).  The ML algorithm is integrated with a 2-bit flash ADC. 
 We synthesize and place the design using low-power 28\,nm CMOS technology and emulate and verify the bitwise performance. The design is expected to operate at less than 300\,$\mu$W with an area of less than 0.2\,mm$^2$.
\end{itemize}

While this study demonstrates significant strides, there are a number of areas where we can improve performance and realism, which we leave to future studies.  We classify them them roughly into algorithmic and microelectronics advancements.  

\paragraph{Algorithms:} One straightforward extension to our studies to improve the data reduction performance of our approach is to train dedicated classification of untracked clusters.  Beyond that, one other very important technique for reducing pixel data is to expand the task to featurize the cluster data after filtering.  We can transform raw cluster data into physics features, such as position and angle, and their related uncertainties.  That will be especially useful in downstream systems to decrease data bandwidth and computations and system complexity by reducing tracking combinatorics.  Algorithmically, we have not yet taken full advantage of the cluster time-evolution which holds additional information.  This can, perhaps, be naturally captured using neuromorphic approaches which can treat the pixels as spikes in an imaging system~\cite{schuman2017neuromorphic,Gallego2022event}.  Neuromorphic computing has been demonstrated as a promising technology for edge applications \cite{schuman2022opportunities}. Spiking Neural Networks (SNN) and their inherent temporal dynamics have been shown to be suitable for processing sparse streaming data and for deployment on low-power neuromorphic platforms closely integrated with the sensors \cite{patton2022neuromorphic,birkoben2020spiking, kosters2023benchmarking}. In other work, we have also demonstrated an initial analysis of the neuromorphic approach for smart-pixels that has yielded models with fewer parameters than in DNNs \cite{kulkarni2023}. Beyond these ideas for improvement, increasing realism will be important in future studies (for example, cases where multiple nearby clusters require arbitration).  

\paragraph{Microelectronics:}  To fully exploit the algorithmic approaches in this work and future work discussed above, we must continue to explore methods for improving computational efficiency.  We mention here a non-exhaustive list of potential approaches for emerging microelectronics technologies~\cite{Deiana:2021niw}. Analog computations are much more efficient than digital approaches for neural networks, and they could be particularly well-suited to sensor data, but are more challenging to design and simulate, and are less reliable.  Novel emerging technologies for efficient neuromorphic compute such as memristors offer promising advances for power and speed as well. Finally, we also mention the potential of 3D stacking as a way to improve computational area without compromising on system constraints.  

It is likely that a combination of these additional ideas is needed to realize a ``smart pixel'' sensor that can meet particle physics experimental needs.  However, having such a sensor would be a transformative technology for particle physics and demonstrating this capability would enable advances in many other scientific domains as well.  

\section*{Acknowledgements}

This work was completed using computing resources at the Fermilab Elastic Analysis Facility (EAF). We thank Burt Holzman for computing support. 

We acknowledge the Fast Machine Learning collective as an open community of multi-domain experts and collaborators. 

We acknowledge the CMS collaboration for their collection of untracked pixel clusters that serve as a valuable background dataset.

DB, JD, GDG, FF, LG, JH, RL, BP, GP, CS and NT are supported by Fermi Research Alliance, LLC under Contract No. DE-AC02-07CH11359 with the Department of Energy (DOE), Office of Science, Office of High Energy
Physics. 
JD, FF, BP, GP, and NT are also supported by the DOE Early Career Research Program. 
NT is also supported by the DOE Office of Science, Office of Advanced Scientific Computing Research under the “Real-time Data Reduction
Codesign at the Extreme Edge for Science” Project (DE-FOA-0002501).  AB is supported through NSF-PHY award 2013007.  MS is supported by NSF-PHY award 2012584.  CM is supported by NSF-PHY award 2208803. KD is supported in part by the Neubauer Family Foundation Program for Assistant Professors and the University of Chicago. AY and SK are supported by the DOE Office of Science Research Program for Microelectronics Codesign (sponsored by ASCR, BES, HEP, NP, and FES) through the Abisko Project. 
MSN is supported through NSF cooperative agreement OAC-2117997, the DOE Office of Science, Office of High Energy Physics, under Contract No. DE-SC0023365, and the Discovery Partners Institute under the "Democratizing AI Hardware with an Open-Source AI-Chip Design Toolkit" Project.

\printbibliography
\clearpage

\appendix

\section{Validation of the dataset}
\label{app:dataset}

This appendix contains additional validation plots of the Smart Pixels dataset~\cite{zenodo}. Note that this dataset contains no corrections for losses due to tracking efficiency at low $p_T$.

\begin{figure}[htbp]
  \centering
  \subfloat[$\cot\alpha$]{\includegraphics[width=0.45\textwidth]{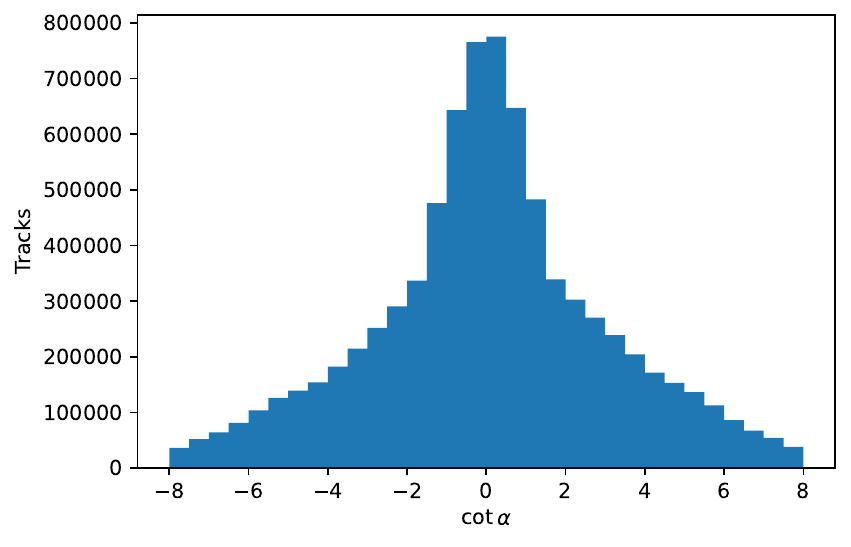}}
  \subfloat[$\cot\beta$]{\includegraphics[width=0.45\textwidth]{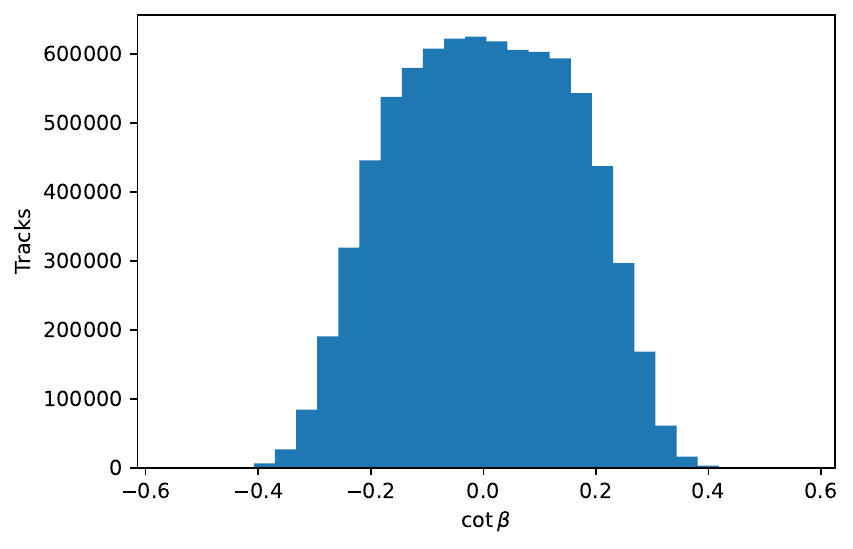}}
  \caption{Simulated particle angle of incidence in the (a) $x-z$ and (b) $y-z$ plane (the bending plane of the magnetic field). }
  \label{angles}
\end{figure}

\begin{figure}[htbp]
  \centering
  \includegraphics[width=0.45\textwidth]{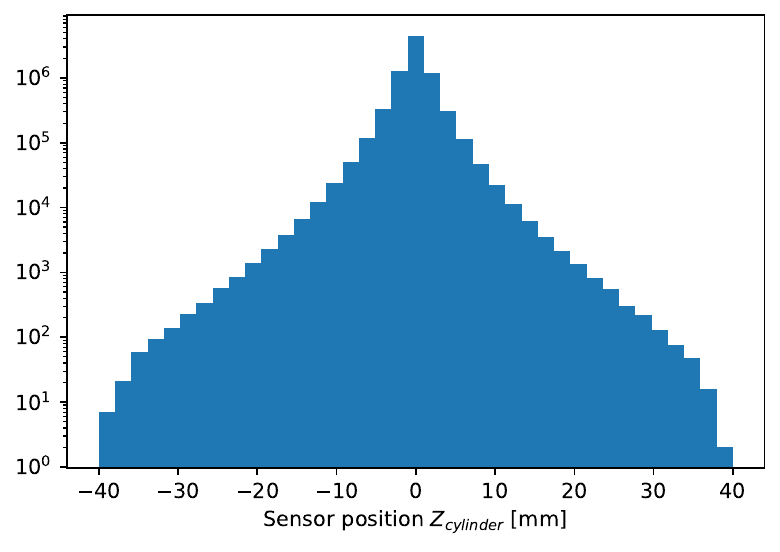}
   \caption{The position of the sensor along the axis of the cylindrical barrel layer, where 0 indicates the particle's point of origin.}
  \label{zcylinder}
\end{figure}

The distribution of simulated incidence angles is shown in Figure \ref{angles}. Larger values of $\cot\alpha$ are achieved compared to $\cot\beta$ due to the position of the sensor at various points along the axis of a cylindrical detector barrel, as shown in Figure \ref{zcylinder}.

\begin{figure}[htbp]
  \centering
  \subfloat[$(x,y)$]{\includegraphics[width=0.45\textwidth]{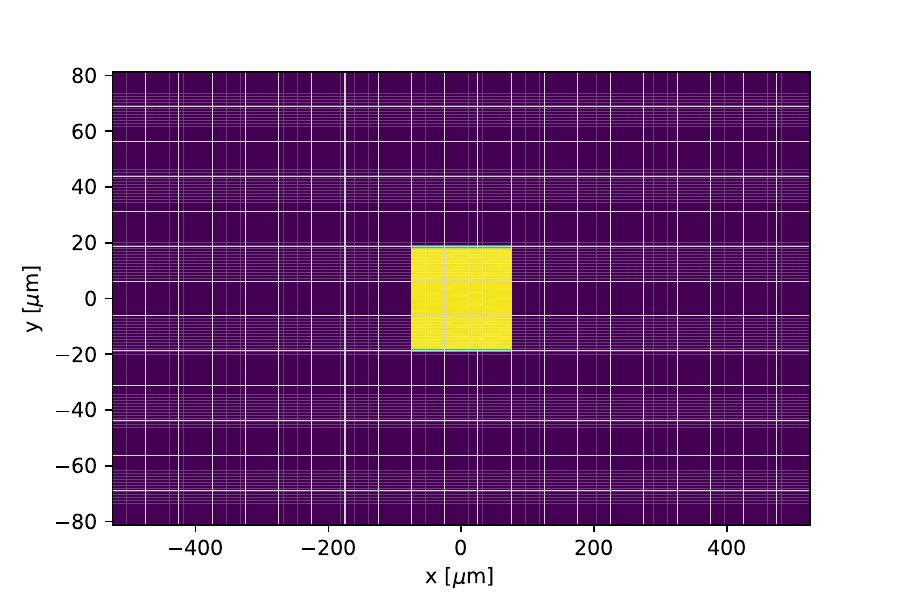}}
  \subfloat[$y_0$]{\includegraphics[width=0.45\textwidth]{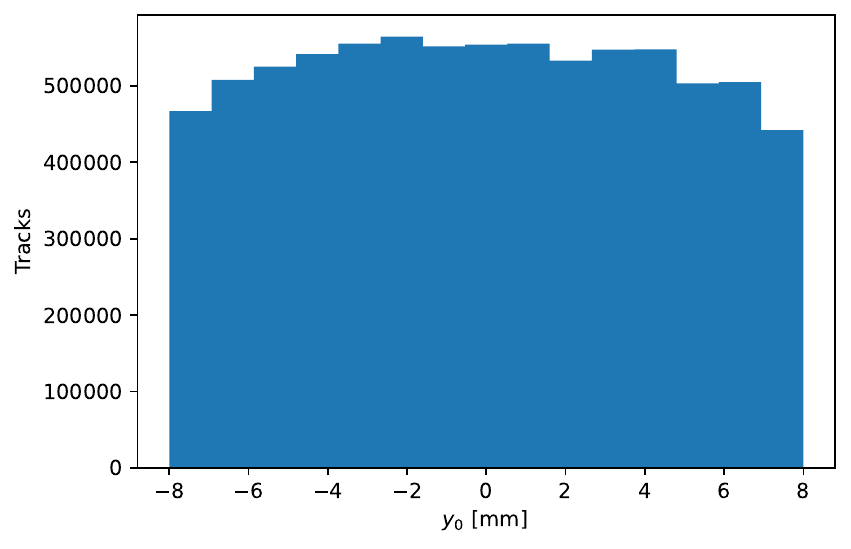}}
  \caption{(a) The position at which the simulated particle crosses the sensor midplane is distributed uniformly across the central $3\times3$ pixel array. (b) The $y_0$ position is approximately uniform across the 1.6cm long sensor.}
  \label{xy}
\end{figure}

The distribution of impact position on the sensor mid-plane is shown in Figure \ref{xy}a, and is uniformly distributed across the central $3\times3$ pixel array.  The distribution of the $y_0$ position of each cluster is shown in Figure \ref{xy}b.

\begin{figure}[htbp]
  \centering
  \subfloat[Positive charge]{\includegraphics[width=0.45\textwidth]{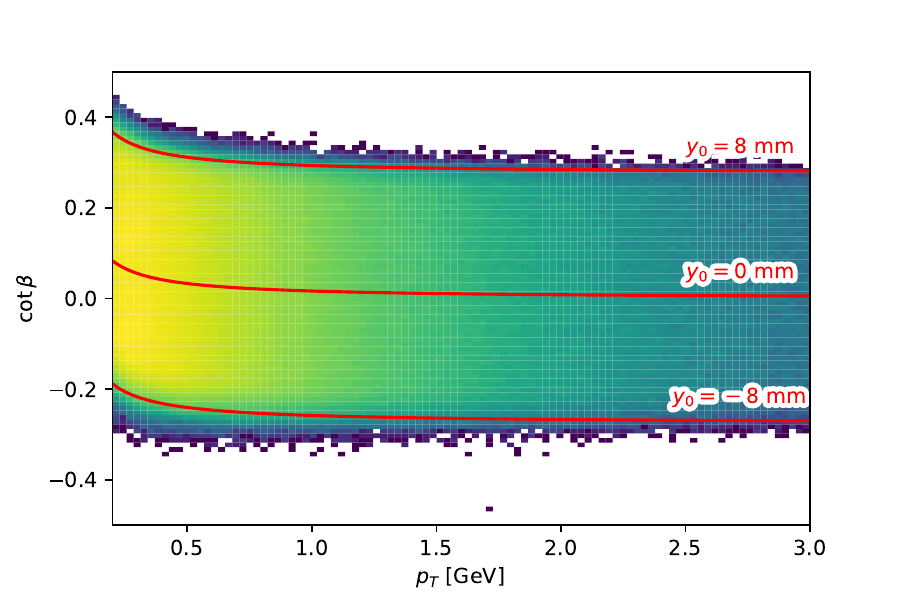}}
  \subfloat[Negative charge]{\includegraphics[width=0.45\textwidth]{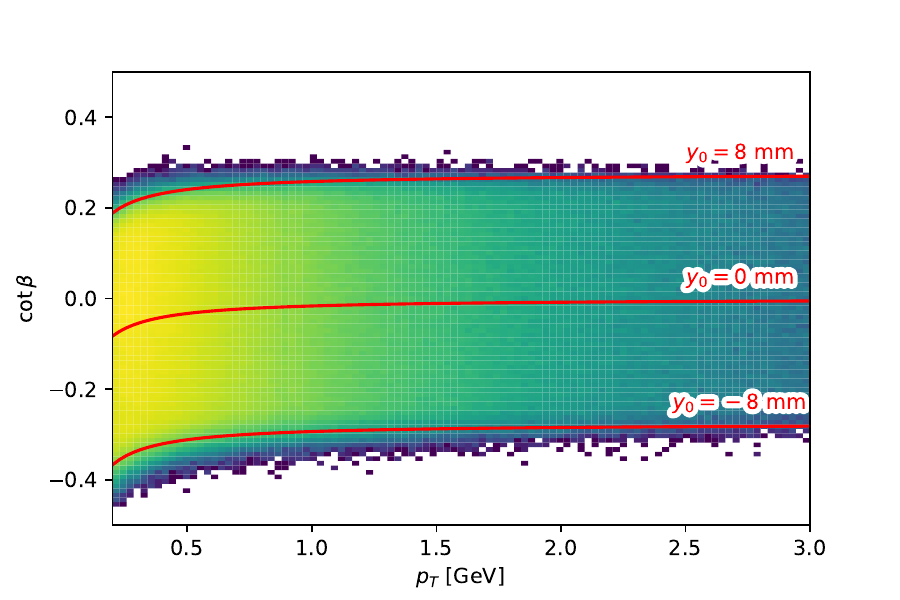}}
   \caption{Relationship between $\cot\beta$ and $p_T$ for particles of different electric charge. Red lines denote the relationship at a fixed $y_0$ position.}
  \label{ptBeta}
\end{figure}

The relationship between the particle $p_T$ and its incident angle in the bending plane of the magnetic field $\beta$ is shown in Figure \ref{ptBeta}. The expected dependence of $\beta$ on $p_T$ and $y_0$ can be derived analytically:
\begin{align}
    \beta = \pi/2 - \Delta\phi - \arctan(y_0/R)
\end{align}
where R denotes the radial position of the sensor. The $p_T$ dependence is embedded in the quantity $\Delta\phi$, which denotes the change in the direction of the charged particle due to the magnetic field:
\begin{align}
    \sin(\Delta\phi) = qRB/(2p_T)\\
\end{align}
The expected curves are overlaid on the simulated data distribution for the central and most extreme values of $y_0$. 

\begin{figure}[htbp]
  \centering
  \subfloat[]{\includegraphics[width=0.44\textwidth]{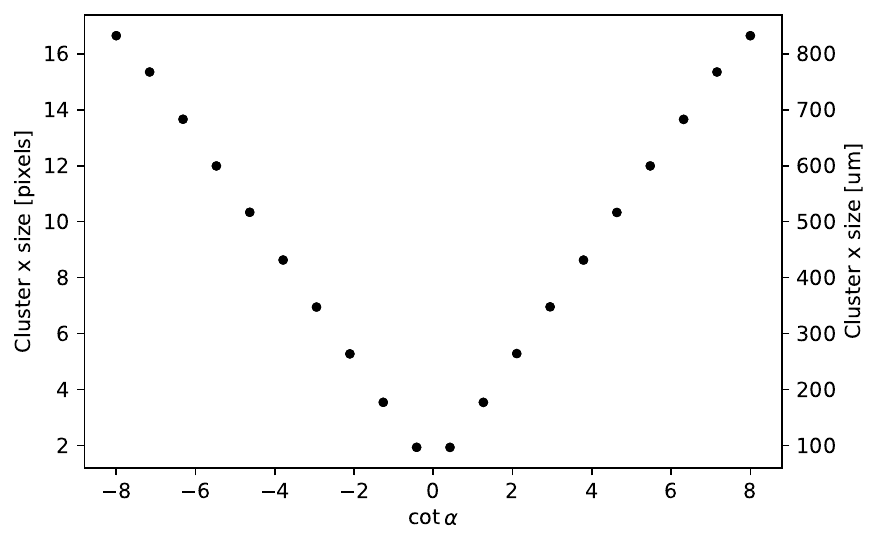}}
  \subfloat[]{\includegraphics[width=0.42\textwidth]{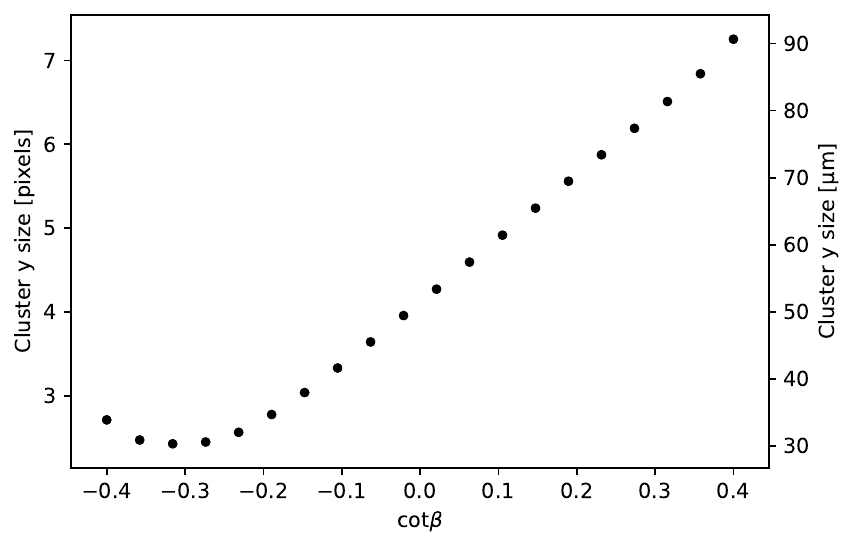}}
   \caption{Dependence of the cluster size on the incident angle. Panel (a) shows the $x$ size vs. angle $\alpha$ in the $x-z$ plane, while (b) shows the $y$ size vs. angle $\beta$ in the $y-z$ plane. In the bending plane of the magnetic field (b), the minimum cluster size occurs at $\cot\beta<0$ due to Lorentz drift.}
  \label{sizeVSangle}
\end{figure}

The relationship between the cluster size and incident angle is shown in Fig\ref{sizeVSangle}a and b for $\alpha$ and $\beta$ respectively. As expected, the cluster size is proportional to the cotangent of the incident angle.

\begin{figure}[htbp!]
  \centering
  \includegraphics[width=0.6\textwidth]{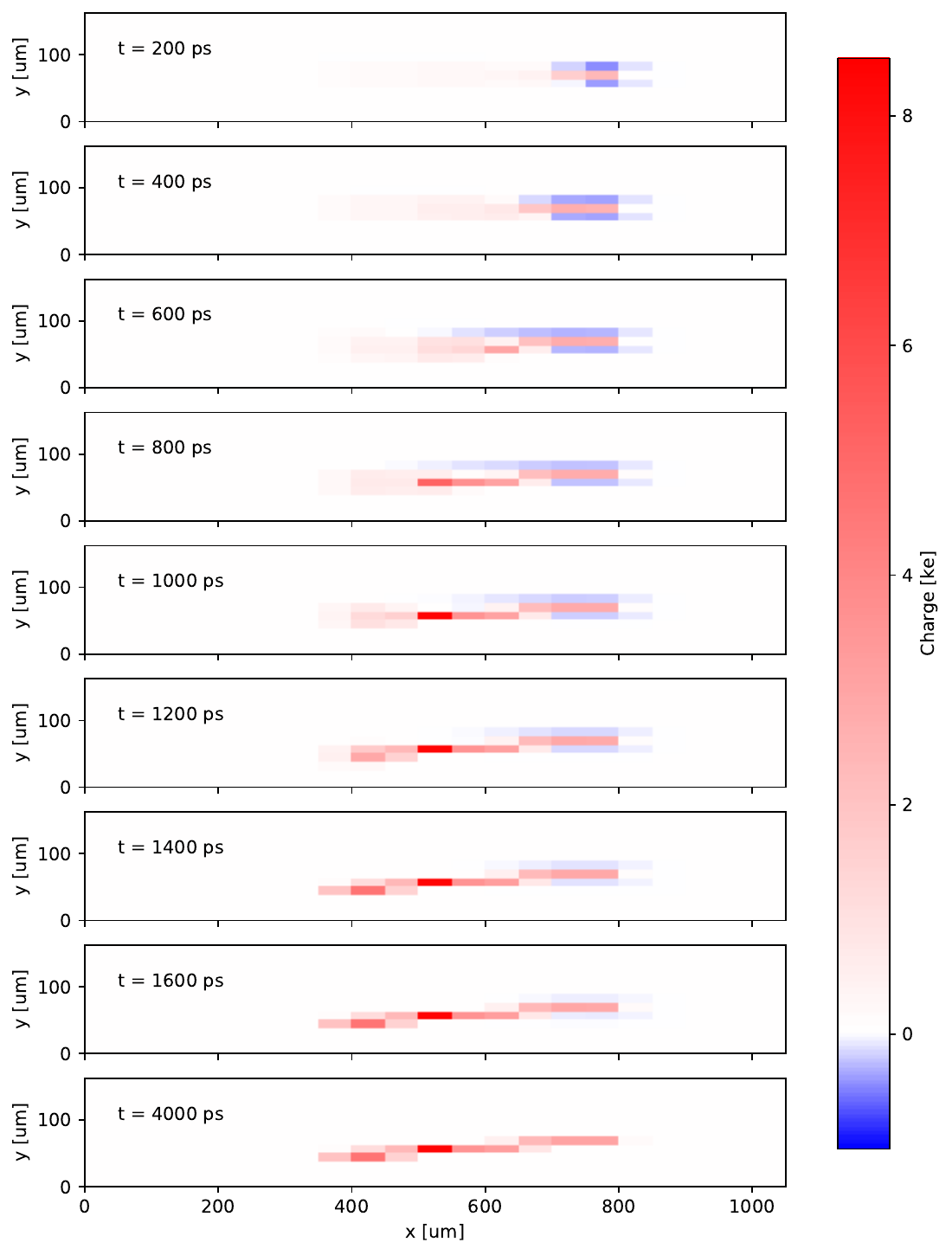}
  \caption{Time evolution of a charge cluster in the simulated $21\times13$ pixel array. The incident particle has $y_0=2.3$ mm and $p_T=1.9$ GeV. The color scale represents the number of electrons collected after the time denoted. Blue indicates induced negative charge in the pixel.}
  \label{timing11}
\end{figure}

\begin{figure}[htbp!]
  \centering
  \includegraphics[width=0.6\textwidth]{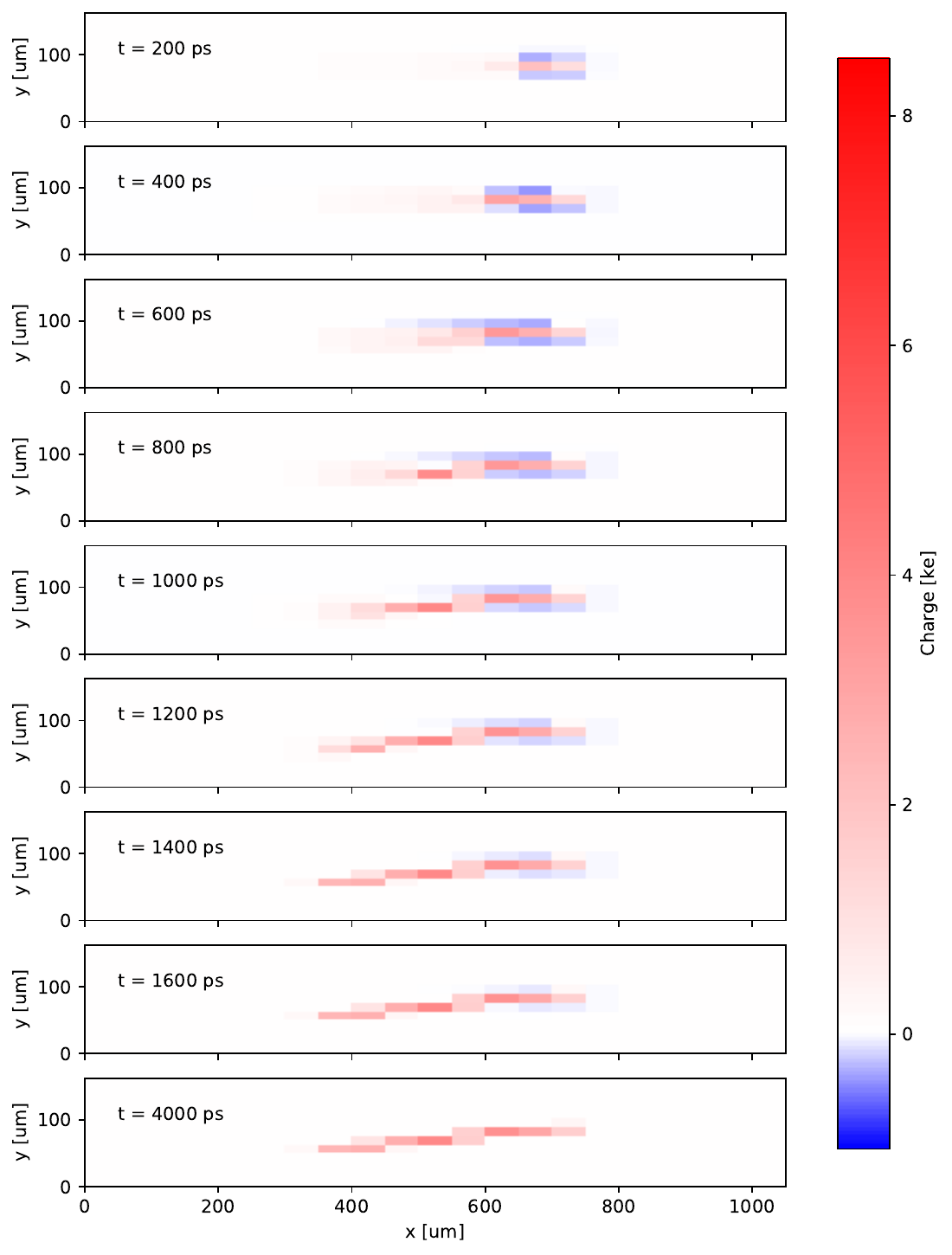}
  \caption{Time evolution of a charge cluster in the simulated $21\times13$ pixel array. The incident particle has $y_0=2.3$ mm and $p_T=135$ MeV. The color scale represents the number of electrons collected after the time denoted. Blue indicates induced negative charge in the pixel.}
  \label{timing40}
\end{figure}

Finally, the time evolution of two example charge clusters is shown in Figures \ref{timing11}-\ref{timing40}. 

\FloatBarrier
\section{Choice of ADC mapping}
\label{app:adc-bits}

\begin{figure}[htbp]
  \centering
  \includegraphics[width=0.65\textwidth]{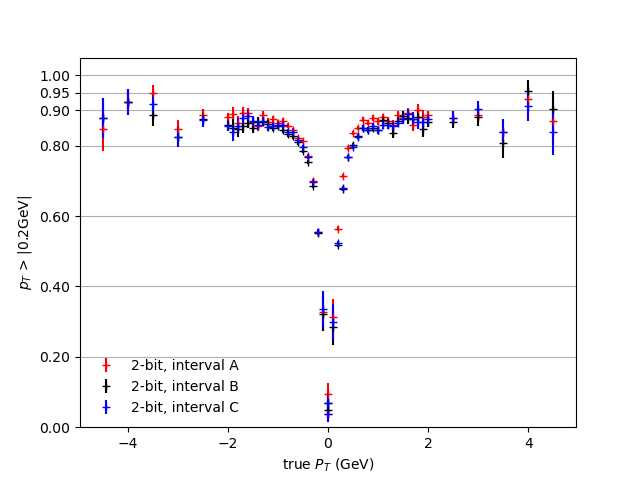}
   \caption{Classifier acceptance vs. track $p_T$ using different 2-bit ADC quantization. Positive and negative values of $p_T$ represent the performance on clusters initiated by particles of positive and negative charge, respectively.}
  \label{fig-adc-2bits}
\end{figure}

The mapping of charge intervals to ADC value was explored for an ADC with 1, 2, 3, and 4 bits.  Three 2-bit ADC quantizations are shown in Table \ref{adc-2bits}. Intervals A correspond to the selected baseline shown in Table~\ref{quantized-inputs}. Intervals B prioritize granularity at low charge, while Intervals C place bin boundaries farther apart than the baseline. Figure \ref{fig-adc-2bits} shows the acceptance vs. $p_T$ of the baseline model trained with each set of intervals.

\begin{table}[!htp]
\begin{center}
\begin{tabular}{llll}
\textbf{ADC output} & \textbf{Intervals A [$e^-$]} & \textbf{Intervals B [$e^-$]} & \textbf{Intervals C [$e^-$]}\\ \hline
00 & $<400$ & $<400$ & $<400$\\
01 & $400-1600$ & $400-800$ & $400-2500$\\
10 & $1600-2400$ &$ 800-1200$ & $2500-5000$\\
11 & $>2400$ & $>1200$ & $>5000$ \\
\hline
\end{tabular}
\caption{Mapping between 2-bit ADC output and collected charge.}
\label{adc-2bits}
\end{center}
\end{table}

\begin{figure}[htbp]
  \centering
  \includegraphics[width=0.65\textwidth]{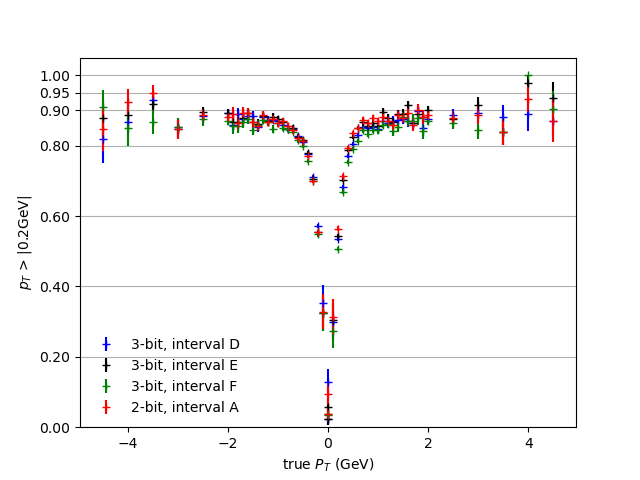}
   \caption{Classifier acceptance vs. track $p_T$ using different 3-bit ADC quantization. Positive and negative values of $p_T$ represent the performance on clusters initiated by particles of positive and negative charge, respectively.}
  \label{fig-adc-3bits}
\end{figure}

The charge intervals studied for 3-bit ADC models are shown in Table \ref{adc-3bits}, and Figure \ref{fig-adc-3bits} shows the acceptance vs. $p_T$ of the baseline model trained with each set of 3-bit intervals.

\begin{table}[!htp]
\begin{center}
\begin{tabular}{llll}
\textbf{ADC output} & \textbf{Intervals D [$e^-$]} & \textbf{Intervals E [$e^-$]} & \textbf{Intervals F [$e^-$]}\\ \hline
000 & $<400$ & $<400$ & $<400$ \\
001 & $400-1600$ & $400-800$ & $400-600$\\
010 & $1600-2400$ & $800-1200$ & $600-800$ \\
011 & $2400-4000$ & $1200-1600$ & $800-1000$\\
100 & $4000-6000$ & $1600-2000$ & $1000-1200$\\
101 & $6000-8000$ & $2000-24000$ & $1200-1400$\\
110 & $8000-10,000$ & $2400-2800$ & $1400-1600$\\
111 & $>10,000$ & $>2800$ & $>1600$ \\
\hline
\end{tabular}
\caption{Example mappings between 3-bit ADC output and collected charge.}
\label{adc-3bits}
\end{center}
\end{table}

For a 4-bit ADC, the following charge intervals were tested:
\begin{itemize}
\item \textbf{Intervals G:} 400 - 3200 e in intervals of 200 e, plus underflow and overflow
\item \textbf{Intervals H:} (high granularity at low charges) 400 - 1800 e in intervals of 100 e, plus underflow and overflow
\item \textbf{Intervals I:} (coarse granularity at low charges) 400-6000 in intervals of 400 e, plus underflow and overflow
\end{itemize}
The acceptance using each set of 4-bit intervals is shown in Figure \ref{fig-adc-2bits} as a function of $p_T$.
The 4 bit quantization did not show a significant increase in performance compared to smaller ADC, perhaps because the intervals between electron thresholds do not scale well with integer addition. 

\begin{figure}[htbp]
  \centering
  \includegraphics[width=0.65\textwidth]{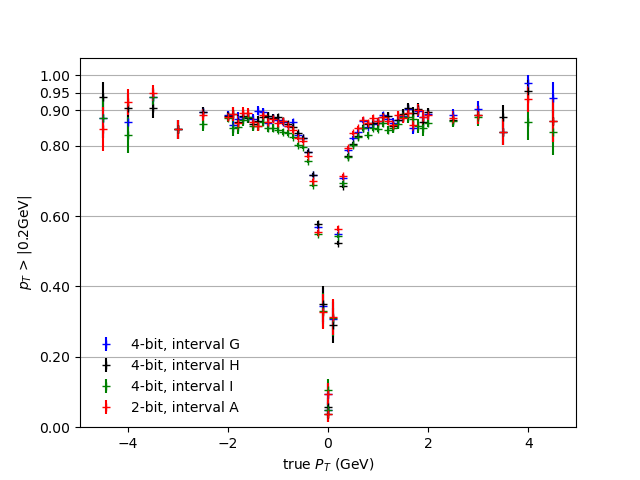}
   \caption{Acceptance vs. track $p_T$ using different 4-bit ADC quantization. Positive and negative values of $p_T$ represent the performance on clusters initiated by particles of positive and negative charge, respectively.}
  \label{fig-adc-4bits}
\end{figure}

\FloatBarrier

\end{document}